\documentclass[fleqn,usenatbib]{mnras}

\usepackage{newtxtext,newtxmath}

\usepackage[T1]{fontenc}
\usepackage{ae,aecompl}

\usepackage{graphicx}	
\usepackage{amsmath}	
\usepackage{amssymb}	

\usepackage{etoolbox}
\makeatletter
\patchcmd\@combinedblfloats{\box\@outputbox}{\unvbox\@outputbox}{}{%
   \errmessage{\noexpand\@combinedblfloats could not be patched}%
}%
 \makeatother

\title[Frequency of Dwarf Multiples]{The Frequency of Dwarf Galaxy Multiples at Low Redshift in SDSS vs. Cosmological Expectations}

\author[G. Besla, et al.]{Gurtina Besla$^{1}$,\thanks{E-mail: gbesla@email.arizona.edu}
David R. Patton$^{2}$,
Sabrina Stierwalt$^{3,4}$
Vicente Rodriguez-Gomez$^{5}$,
\newauthor
Ekta Patel$^{1}$,
Nitya J. Kallivaylil$^{3}$,
Kelsey E. Johnson$^{3}$,
Sarah Pearson$^{6}$,
George C. Privon$^{7}$,
\newauthor
and Mary E. Putman$^{6}$ 
\\
$^{1}$ Steward Observatory, University of Arizona, 933 North Cherry Avenue, Tucson, AZ, 85721, USA\\
$^{2}$ Department of Physics \& Astronomy, Trent University, 1600 West Bank Drive, Peterborough, Ontario, K9L 0G2, Canada \\
$^{3}$Department of Astronomy, University of Virginia, 530 McCormick Road, Charlottesville, Virginia 22904, USA \\
$^{4}$ California Institute of Technology, 1200 East California Boulevard, Pasadena, CA 91125, USA \\
$^{5}$ Department of Physics and Astronomy, Johns Hopkins University, 3400 N. Charles Street, Baltimore, MD 21218, USA \\
$^{6}$Department of Astronomy, Columbia University, Mail Code 5246, 550 West 120th Street, New York, New York 10027, USA\\
$^{7}$ Department of Astronomy, University of Florida, 211 Bryant Space Science Center, Gainesville, FL 32611, USA\\
}

\pubyear{2017}

\begin{document}
\label{firstpage}
\pagerange{\pageref{firstpage}--\pageref{lastpage}}
\maketitle

\begin{abstract}
We quantify the frequency of companions of low redshift ($0.013 < z < 0.0252$), dwarf galaxies ($2 \times 10^8$ M$_\odot <$ M$_{\rm star} < 5 \times 10^9$ M$_\odot$) that are isolated from more massive galaxies in {\it SDSS}
and compare against cosmological expectations using
mock observations of the {\it Illustris} simulation. 
Dwarf multiples are defined as 2 or more dwarfs that have angular separations $> 55\arcsec$, projected separations r$_p < 150$ kpc and relative line-of-sight velocities $\Delta V_{\rm LOS} < 150$ km/s.   While the mock catalogs predict a factor of 2 more isolated dwarfs than observed in {\it SDSS}, the mean number of observed companions per dwarf is $N_c \sim 0.04$, in good agreement with {\it Illustris} when accounting for {\it SDSS} sensitivity limits. 
Removing these limits in the mock catalogs predicts $N_c\sim 0.06$ for future surveys (LSST, DESI), which will be complete to M$_{\rm star} = 2\times 10^8$ M$_\odot$.
The 3D separations of mock dwarf multiples reveal a
contamination fraction of $\sim$40\% in observations from projection effects. 
Most isolated multiples are pairs; triples are rare and
it is cosmologically improbable that bound groups of dwarfs with more than 3 members exist within the parameter range probed in this study.
We find that $<$1\% of LMC-analogs in the field have an SMC-analog companion.
The fraction of dwarf ``Major Pairs'' (stellar mass ratio $>$1:4) 
steadily increases with decreasing Primary stellar mass, whereas the cosmological ``Major Merger rate'' (per Gyr) has the opposite behaviour.
We conclude that 
cosmological simulations can be reliably used to constrain the fraction of dwarf mergers across cosmic time.

\end{abstract}




\section{Introduction}

Low mass, dwarf galaxies (M$_{\rm star} < 5 \times 10^9$ M$_\odot$) are 
the most common class of galaxy at all redshifts
\citep{Binggeli98,Karachentsev13}, and yet, their pair/group fractions have not 
been quantified observationally and compared to cosmological expectations in a consistent manner. 
In contrast, the interaction frequency between massive galaxies 
is actively studied, both observationally \citep[e.g.,][ etc.]{Zepf89,Conselice03,Bundy04,Bell06,Lin08,Carlberg00,Patton02,Patton08,Lotz11,Patton16,Man16,Mundy17,Ventou17,Mantha18}
and theoretically \citep[][etc.]{Blumenthal84, Lacey93, Berrier06, Maller06,Guo08, Hopkins10, Rodriguez15}.  
Consequently, little is known about the frequency of 
dwarf-dwarf galaxy interactions and their role in the evolution of low mass galaxies. 

In this study, we quantify the fraction of dwarf galaxies with low mass companions at
low redshift utilizing both observational data
from the Sloan Digital Sky Survey ({\it SDSS}) and ``mock'' galaxy catalogs
created using the {\it Illustris-1} cosmological simulation, 
\citep{Vogelsberger14, Nelson15}. 
Our goal is to establish the reliability with which cosmological simulations can trace the hierarchical processes that are expected to influence the evolution of dwarf galaxies.

Environmental effects  
(the tidal field of a massive host, ram pressure stripping, etc.) are often presented as the dominant drivers of dwarf galaxy evolution. This is motivated by the fact that non-star forming, 
gas poor (quenched) dwarf galaxies are almost exclusively found 
in proximity to a massive host \citep{Geha12, Sanchez13, Stierwalt15, Bradford15}.  Satellite 
dwarf galaxies of the Local Group \citep{vandenB06, Grcevich09, Spekkens14} and the Local Volume \citep{Weisz11}
all exhibit similar distance-morphology relationships. 
Consequently, the role of mergers or interactions between dwarfs {\it prior} to their capture as a satellite of a massive galaxy
 has been largely ignored.  However, this picture is rapidly changing.
 
While environmental processes are clearly needed to quench dwarf galaxies \citep[e.g.,][]{Peng10}, they are 
not the only factors governing the evolution of low mass galaxies. High-resolution dark matter simulations of Milky Way type halos reveal that
10\% of surviving dwarf-mass satellites have experienced a major merger since z=1 \citep{Deason15} and that 10\% of dwarfs host a satellite of at least 10\% its mass at z=0 \citep{Sales13}.
Furthermore, 30-60\% of all surviving dwarf satellites are expected to have 
been accreted as part of a low mass group \citep{Wetzel15}. Cosmologically, the tidal
pre-processing of dwarf galaxies in low mass multiples is not expected to be a rare 
phenomenon across cosmic time and presents an alternative pathway for galaxy evolution at the low mass end. However, these theoretical predictions have yet to be tested systematically against observations. 

We are only just starting to understand the impact of galactic pre-processing on 
the star formation histories (SFHs) and baryon cycle (supply and removal of gas) of dwarf galaxies. Earlier studies have suggested that many star bursting dwarf galaxies have companions \citep{Noeske01}.
Recently, the TiNy Titans ({\it TNT}) Survey studied 
the star formation rates (SFRs) and gas content 
of dwarf galaxy multiples (pairs and groups with stellar masses of 
$10^7$ M$_\odot <$ M$_{\rm star} < 5\times10^9$ M$_\odot$) 
identified at low redshift ($0.005 < z < 0.07$) in 
{\it SDSS} \citep[{\it TNT};][]{Stierwalt15} and in the Local Volume \citep[$<$ 30 Mpc, {\it TNT-LV};][]{Pearson16}.
The {\it TNT} and {\it TNT-LV} surveys are 
designed to study the dwarf-dwarf merger sequence and its importance to the 
evolution of low mass galaxies. This paper represents the theoretical counterpart to these studies.

The {\it TNT} surveys have provided  
mounting evidence that interactions between low mass galaxies
do impact the structure and SFHs of dwarf galaxies \citep[e.g.,][]{Privon17}.
In particular, \citet{Stierwalt15} found that SFRs are elevated in dwarf pairs 
relative to their non-paired counterparts;  dwarf-dwarf interactions appear to drive enhanced
star formation. Furthermore, \citet{Pearson16} illustrate that 
dwarf pairs in the Local Volume are often found with 
extended gaseous envelopes, compared to their non-paired counterparts, suggesting
that the efficiency of gas removal can also be increased through dwarf-dwarf interactions. This is true for the Magellanic System, the closest example of a dwarf-dwarf interaction \citep{Putman03, Nidever10, Besla10,Besla12,Besla13,Diaz11,Guglielmo14}.  
Indeed, \citet{Marasco16} use the EAGLE cosmological simulation \citep{Schaye15} to illustrate that satellite-satellite encounters may be more important 
than environmental effects (tidal/ram pressure stripping by the host) in the gaseous evolution of dwarf satellite galaxies.

With next generation photometric and spectroscopic instruments, such as LSST, DESI, WFIRST, etc.,  
new detections of low mass (or low surface brightness) galaxies with companions are forthcoming. 
A glimpse of what lies ahead is highlighted by the recent
discovery of seven isolated 
groups of dwarfs with 3-5 members in the {\it TNT} sample \citep{Stierwalt17}. 
The newly discovered {\it TNT} groups provide us with a window into a 
process of hierarchical evolution that 
may have been much more common at high redshift. However, there is no existing framework to understand 
how these low redshift groups fit in the prevailing cosmological model.  

The average satellite mass functions of dwarf galaxies have been quantified in cosmological simulations \citep{Dooley17}
 and compared to observations in {\it SDSS} \citep{Sales13}.  
Here, we expand such methods to create mock catalogs of cosmological dwarf multiples (pairs/groups) in isolated environments, accounting for the sensitivity limits of {\it SDSS} and line-of-sight properties (projected separation and relative velocities; including peculiar motions) that control the observational definition of companionship.  Our goal is to assess whether or not the observed and theoretical fraction of dwarf multiples agree within the redshift and mass ranges where they can be reasonably compared: stellar masses between 
$2\times 10^8 $ M$_\odot$ < M$_{\rm star} < 5\times  10^9$ M$_\odot$ at low redshift (Volume $\lesssim$ (100 Mpc)$^3$; $0.013 <$ z $<$ 0.0252).
Specifically, we aim to address the following questions:

\begin{enumerate}
\item What is the observed fraction of dwarfs in a pair or group vs. cosmological expectations?
\item What is the contamination fraction of dwarf multiples owing to projection effects?
\item What do cosmological simulations predict for the frequency of dwarf multiples in the era of deep photometric surveys like LSST? 
\item What is the z$\sim$0 fraction of dwarf ``Major Pairs'' (stellar mass ratio $>1:4$)? 
\item What is the observed frequency of Magellanic Cloud analogs in the field vs. cosmological expectations? 
\item Are the recently-discovered {\it TNT} dwarf groups \citep{Stierwalt17} consistent with cosmological expectations?  
\end{enumerate}

 Through this analysis we will establish the reliability of such simulations as probes of hierarchical processes at low masses at z$\sim$0, lending credibility to their usage as predictors of the frequency of dwarf-dwarf interactions and mergers across cosmic time.

In Section~\ref{sec:GalSec} we describe our methodology to assign properties to galaxies (stellar mass, redshifts, isolation) and define survey volumes in both {\it SDSS} and {\it Illustris}.  
In Section~\ref{sec:NCount} we quantify number counts of all galaxies (dwarfs and massive galaxies) and isolated dwarfs. 
Results for the frequency of dwarf multiples in {\it SDSS} and {\it Illustris} and predictions for next generation surveys are summarized in Section~\ref{sec:FreqPairs}. We discuss the dwarf ``Major Pair'' fraction, the frequency of Magellanic Cloud analogs and place the {\it TNT} groups of dwarfs in a cosmological context in Section~\ref{sec:Discussion}. Finally, we conclude in Section~\ref{sec:Conclusions}.

\section{Galaxy Catalogs}
\label{sec:GalSec}

We define a ``dwarf galaxy'' as a galaxy with a stellar mass of $2\times 10^8$  M$_\odot <$  M$_{\rm star} < 5 \times 10^9$ M$_\odot$. The upper mass limit corresponds to systems slightly more 
massive than the LMC 
\citep[M$_{\rm star} = 3 \times 10^9$ M$_\odot$;][]{van02}. 
According to the Tully-Fisher relation, this definition generally 
excludes galaxies with rotation curves that peak at 100 km/s or larger \citep{Lelli14}.
For reference, the LMC's rotation curve peaks at $\sim$90 km/s \citep{van14}. 

The lower mass limit corresponds to the stellar mass of the 
SMC \citep[M$_{\rm star} \sim 2 \times 10^8$ M$_\odot$;][]{Stanimirovic04, van09}.
The completeness of the SDSS catalog drops rapidly with decreasing stellar mass as a function of redshift. Galaxies of  stellar mass of $\sim 2\times 10^8$ M$_\odot$ are the lowest mass galaxies that are complete in SDSS at the lowest redshift considered in this study (z=0.013; see Section~\ref{sec:Volume}).

Galaxies with stellar masses larger than $5 \times 10^9$ M$_\odot$
are referred to as ``Massive Galaxies''. 
Dwarf multiples (pairs and groups) do not 
survive for very long as bound configurations about massive galaxies \citep[e.g.,][]{Gonzalez16}.  
We thus require our dwarf galaxy sample to be sufficiently isolated from such systems, as 
described in Section~\ref{sec:Isol}.

In the following, we describe how stellar masses are defined for all galaxies in both 
the observational and cosmological data sets.  Note that the cosmological galaxy samples will be 
referred to as ``mock'' galaxy catalogs.

\subsection{Observational Galaxy Catalog: \emph{SDSS}} 
\label{sec:ObsSample}

Following \citet{Stierwalt15}, our observational sample is drawn from the Legacy area of the 
{\it SDSS} Data Release 7 spectroscopic catalog \citep{Abazajian09}. We utilize only the 
continuous footprint, which covers 7296 deg$^2$ of the sky.
Spectroscopic completeness of galaxies imaged is estimated at 88\% for galaxies with 
$14.5 < m_r < 17.5$ \citep{Patton08}. Note that low 
surface brightness galaxies and very close pairs will be preferentially missed.  
We apply lower limits on the anglar separation 
between dwarf galaxies (r$_p > 55\arcsec$; i.e. projected distances of 15-25 kpc depending on the redshift) to avoid fiber collisions that can miss close pairs (Section~\ref{sec:DwarfCompanion}).

We select galaxies from the {\it SDSS} value-added catalog of \citet{Simard11}, which is a 
reprocessing of the {\it SDSS} photometry using bulge-disk decomposition and an improved 
handling of de-blending in crowded systems, such as close galaxy pairs. Stellar masses
and associated errors are taken from \citet[][hereafter M14]{Mendel14}. We use the Sersic {\it ugriz} total 
stellar mass fits of M14, as recommended in their Appendix B.2.1 
\citep[see also][]{Patton13}.  

We assume Gaussian errors and
sample the M14 stellar mass errors randomly to generate 500 unique realizations of the entire stellar mass catalog, each with a different set of stellar masses allowed within the errors.
All statistics presented in this study are computed as the mean and standard deviation over these
500\footnote{We have also repeated this study using 1000 unique realizations and found similar results within 1$\sigma$.} realizations.

The M14 catalog is restricted to 
galaxies with $14 < m_r < 17.77$, where $m_r$ is the extinction-corrected
$r$-band Petrosian magnitude from the {\it SDSS} database. As such, we 
supplement the {\it SDSS} Massive 
Galaxy sample with bright galaxies ($m_r < 14$) 
identified in the NASA-Sloan Atlas (NSA) and adopt the associated stellar masses and redshifts 
quoted in the catalog\footnote{http://www.nsatlas.org/data}.  We do not randomly sample the stellar mass errors for the 
massive galaxy sample as we are not interested in the intrinsic properties of these galaxies. 

Redshifts are adopted for dwarfs ($m_r > 14$) from the {\it SDSS} data release 9 (DR9)
spectroscopic catalog,
which has an average redshift error of order $10^{-5}$. In contrast, the average redshift error
in DR7 is $10^{-3}$, which is too large to
meaningfully extract kinematic information for the
dwarf multiples. 
The corresponding DR9 velocity errors are $\sim$ 4 km/s, computed at the average 
redshift of our sample (z $\sim$ 0.021). 

We assign redshifts to each galaxy in the sample in a similar fashion to the assignment of
stellar mass: we assume Gaussian errors that are randomly 
sampled to generate the 500 unique realizations of the catalog. As such 
redshift errors, although small, are accounted for
 in the determination of both the relative velocities and positions of dwarf multiples. 

There is 98.9\% overlap between galaxies
identified in the \citet{Simard11} catalogs and those in 
DR9 (for z$<$0.05).
However, because
we are using stellar masses from M14, 
who adopt DR7 redshifts, we further require that 
the difference between the DR7 and DR9 redshifts
corresponds to a velocity difference
less than 100 km/s. With this 
requirement, there is a $\sim$97.5\% overlap between the 
DR9 and M14 catalogs.  The remaining $\sim$2.5\% 
of the sample are assigned
DR7 redshifts and errors. 
A velocity error of 100 km/s corresponds to 
a stellar mass error of 0.06 dex at z=0.005 and 
0.01 dex at z=0.0252. This is well below the
average mass error in M14 of 0.1 dex.  
Moreover, the average redshift 
difference between our galaxy samples in DR7 
and DR9 corresponds
to a velocity difference of $\sim$ 15 km/s.
This error is much lower than the velocity constraints
that we later apply to define dwarf multiples 
($\Delta V_{\rm LOS}$ = 150 km/s; see Section~\ref{sec:DwarfCompanion}).

\subsection{Mock Galaxy Catalog: {\it Illustris} \& Abundance Matching}

We utilize data from the {\it Illustris} Project \citep{Vogelsberger14}:
an N-body and hydrodynamic simulation spanning a cosmological volume of (106.5 Mpc)$^3$ to 
derive ``mock'' galaxy catalogs, which will be the basis for comparison to expectations  
from $\Lambda$CDM theory.

In this analysis, we use the highest resolution hydrodynamic {\it Illustris-1} (hereafter {\it Illustris Hydro})
version of the simulation suite, which simulates the growth of structure from z=127 to z=0. 
We have tested all of our results using the dark matter-only {\it Illustris-1-Dark} simulation, and find very good agreement (see Appendix~\ref{sec:DarkCompare}). 

The 
{\it Illustris} Project adopts cosmological parameters 
consistent with WMAP-9 \citep{Hinshaw13}: $\Omega_m$=0.2726, $\Omega_b$=0.0456,  
$\Omega_\Lambda$=0.7274, $\sigma_8=0.809$, $n_s=0.963$ and $h=0.704$.

For this study we will not use stellar masses that result from the explicit star 
formation prescription adopted in  
 {\it Illustris Hydro}.
There are large uncertainties in the star formation prescriptions 
appropriate for low mass galaxies, which can strongly affect the evolution of low mass systems. 
In particular, it has been found that stellar masses of dwarfs in {\it Illustris Hydro} are too high compared
to observations \citep{Genel14}. 
Different subgrid prescriptions for the physics of the ISM are adopted by many teams and no consensus
 has yet been reached by the community.  
Instead, we derive all statistics using {\bf only the dark matter} component of 
subhaloes identified in
{\it Illustris Hydro}. We assign stellar masses via abundance matching utilizing the global baryonic fraction to convert the dark matter component to a total subhalo mass.
  
 We utilize {\it Illustris Hydro} in this analysis
  because the growth of dark matter subhaloes 
is known to be affected by baryonic processes, such as feedback \citep[e.g.][]{Wetzel16}.
As a result, many low mass subhaloes that exist in {\it Illustris-Dark-1} can be destroyed at early times in 
{\it Illustris Hydro} \citep{Chua17}.

Dark matter haloes and their associated substructures (or subhaloes) are identified in {\it Illustris Hydro} 
using the SUBFIND 
halo-finding routine \citep{Springel01, Dolag09}.
Theoretical dwarf and massive galaxy analog samples are then selected from the publicly available SUBFIND subhalo group catalogs generated at the z=0 snapshot
of the simulations.
{\it Illustris Hydro} reaches a dark matter particle mass resolution of m$_{\rm DM}$ = $6.3 \times 10^6$  M$_\odot$ per particle. 
 We place a lower limit of M$_{\rm DM}$ = $5\times 10^9$ M$_\odot$, ensuring  
that all haloes are well-resolved into $>$790 particles. This mass is also roughly the
dynamical mass of the SMC within 3 kpc \citep{Harris06}.


\subsubsection{Assigning Stellar Mass to the Mock Galaxy Catalog}
\label{sec:MockMass}

We assume that the stellar mass associated with a given subhalo is a function of the 
maximum dark matter mass a subhalo has ever attained \citep[see, e.g.][]{Boylan-Kolchin11}.
Hydrodynamic, cosmological simulations have supported this assertion, finding that 
 halo mass grows in unison with stellar mass \citep{Wellons17}.
 We utilize the abundance matching relations in \citet{Moster13} 
to assign 
a stellar mass to each subhalo, as outlined below.   

We first identify the maximal dark matter mass ever
achieved by each subhalo over its cosmic history ($M_{\rm DM,max}$) using  
 merger trees created with the recently developed SUBLINK code 
\citep{Rodriguez15}.  
Because we are searching for dwarf subhalos in close proximity to each other, 
their dark matter distributions are likely truncated by tidal effects, making mass 
assignment via abundance matching less reliable if we choose their final descendent
dark matter masses. 

In addition, we 
require subhaloes to be separated by at least 5 times the gravitational softening length, 
which helps to
 assign reliable halo masses to merging systems.   
 Similar methods have been applied to select 
Magellanic Cloud analogues in the Millennium-II simulation \citep{Boylan-Kolchin11}. 

A stellar mass is then assigned to a subhalo using the z=0 \citet{Moster13} relations (their equations 11-14), and the maximal total subhalo mass, $M_{\rm subhalo} = M_{\rm DM,max} \times \Omega_m/(\Omega_m - \Omega_b)$.
To account for the scatter in the halo mass-stellar mass relation, we assume Gaussian errors using the 1$\sigma$ errors 
for each parameter, as listed in table 1 of \citet{Moster13},
 to randomly assign each subhalo a stellar mass. This procedure of assigning stellar mass 
is repeated every time we select a set of mock galaxies (see Section~\ref{sec:Volume}).

\begin{figure}
\begin{center}
\mbox{\includegraphics[width=3.8in]{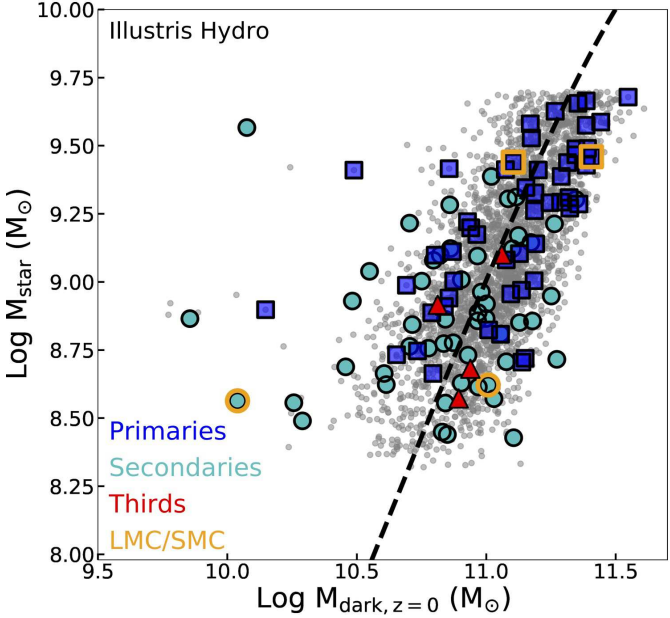}}
\end{center}
 \caption{ \label{fig:DMStars} 
The stellar mass 
 is plotted vs. the descendent dark matter 
mass of each mock dwarf in the {\it Illustris Hydro} simulation at z=0. 
All mock dwarfs are isolated from more massive galaxies (see Section~\ref{sec:Isol}).  
Gaussian errors are assumed using 1$\sigma$ errors 
for each parameter in the \citet{Moster13} relation
 to randomly assign a stellar mass to each subhalo, using the maximal mass the subhalo achieves.  Isolated multiples (groups and pairs) are indicated by larger symbols, whereas companionless dwarfs are marked by grey points. 
Isolated analogues of the LMC+SMC pair (see Section~\ref{sec:LMC}) are marked by orange circles (SMC) and squares (LMC).
The dashed black line indicates the mean abundance matching relation at z=0 from \citet{Moster13}. 
Simulation data are presented for one representative sightline through the simulation volume.
In practice this process is repeated
for 500 different sightlines through the 
simulation volume, resampling the abundance matching relation errors
each time (see Section~\ref{sec:Volume}).
}
 \end{figure}

The resulting halo-stellar mass pairing will scatter about the mean 
abundance matching relation; see Fig.~\ref{fig:DMStars}, where the dashed black line illustrates the mean relation from \citet{Moster13}.
For a halo of mass 10$^{11}$ M$_\odot$, abundance matching yields a mean stellar mass of $\log$(M$_{\rm star})$ = 9.0$\pm$ 0.3. 
Uncertainties on stellar mass for the mock catalog are thus roughly a factor 3-4 larger than those in the M14 catalog.  

Note that there are a number of cases in Fig.~\ref{fig:DMStars} where the derived stellar mass is larger than
the scatter about the abundance matching relations at z=0 (points to the far left). 
This is because $M_{\rm DM, max}$, which was used to assign the 
stellar mass, is larger than $M_{\rm DM, z=0}$. In each of these 
cases the dwarf subhalo has a companion dwarf in close proximity r$_p < 30$ kpc. However, in some cases the dwarf companion is at smaller angular separations than 55$\arcsec$ and so we do not count it as part of a multiple (marked as grey points instead). Because the dwarf subhalos are close together, their individual dark matter masses are ill-defined: either tides have stripped the outer halo, or the halos overlap sufficiently that most of the mass is assigned to the companion, causing the dark matter mass to decrease substantially at z=0 relative to their maximal halo mass. This is the reason the maximal halo mass is utilized to assign stellar mass.

\subsubsection{Assigning Redshifts to the Mock Galaxy Catalog}

In this section, we describe how redshifts and line of sight velocities are assigned to our mock dwarfs. This methodology is complementary to the studies of \citet{Snyder17} and \citet{Behroozi15}, who utilize the {\it Illustris} and {\it Bolshoi} simulations, respectively, 
 to create mock galaxy catalogs and quantify the pair fraction of massive galaxies at higher redshift in comparison to observations.

Here, redshifts are assigned to mock dwarfs using 
the 3D distance and $v_{\rm los}$ of each subhalo in the simulation 
volume with respect to an observer placed randomly in the box.  The redshift is computed as a
combination of the redshift from the Hubble Flow, $z_H$, and the peculiar motion of the subhalo ($v_{\rm LOS}$)
along the observer's line-of-sight (red or blue shift).
The observed redshift for the mock galaxy 
is thus
\begin{equation}
z =  z_H + \frac{v_{\rm LOS}}{c_{\rm light}}
\end{equation}
\noindent where $v_{\rm LOS}$ can be positive or negative. $z_H$ is determined by first computing the 
3D distance of the mock galaxy to the observer (D$_{\rm LOS}$). 
Then we assign the cosmological redshift using a look-up table of comoving distances at a given redshift, $D_C(z)$:  
\begin{equation}
D_C(z) = \frac{D_L(z)}{1+z} 
\end{equation}

\noindent where $D_L(z)$ is the luminosity distance computed at a given redshift, $z$, using the {\it Illustris} cosmology. The cosmological redshift of the mock galaxy, $z_H$, is defined where $D_{\rm LOS} = D_C(z_H)$.

\subsection{Selection of Isolated Dwarf Galaxies}
\label{sec:Isol}

Low mass galaxy groups are unlikely to survive as a bound configuration 
when in proximity to a massive galaxy 
\citep{Gonzalez16}. The tidal field of the host can disrupt groups of dwarfs after even one 
pericentric passage.  This explains 
 the rarity of LMC+SMC binaries about Milky Way type hosts as determined by both observational 
 \citep{Liu11,Robotham12} and cosmological surveys \citep{Boylan-Kolchin11, Busha11, Gonzalez13}
 and by numerical simulations of the LMC+SMC binary evolution \citep{Besla12, Kalli13}. 
There are indications that the number of dwarf galaxy pairs may be higher in the Local Group \citep{Fattahi13}.
However, the number density of dwarfs
increases substantially in the presence of massive galaxies, increasing the 
likelihood of chance projections mimicking true multiples.   

An isolation criterion also affords us the best comparison sample between the observations 
and simulation data as  
environmental effects can suppress star formation. This can cause extreme discrepancies 
between halo-stellar mass correlations and 
makes dwarfs redder in color and thus harder to detect in surveys like the {\it SDSS}.

We define a dwarf galaxy to be {\bf isolated}  {\it if no Massive Galaxy (M$_{\rm star} > 5\times 10^9$ M$_\odot$) can be identified that satisfies both of the following criteria}:

\begin{enumerate}
\item  A relative line of sight velocity $\Delta V_{\rm LOS} <$ 1000 km/s. \citet{Patton00} demonstrated that associations with massive galaxies at lower velocity separations are unlikely to be random.

\item  Tidal Index, $\Theta > -11.5$. Following \citet{Karachentsev13}, we define the Tidal Index as, 
\begin{equation}
    \Theta =  \log \Bigg( \frac{ {\rm M}_{\rm star} (10^{11} M_\odot)}{r_p^3({\rm Mpc})^3} \Bigg) - 10.97 
    \label{eqn:Tidal}
\end{equation}

\noindent  where M$_{\rm star}$ is the stellar mass of a given massive galaxy in units of $10^{11}$ M$_\odot$ and $r_p$ is the projected separation between the Massive Galaxy and the dwarf \citep[see also][]{Karachentsev04}. \citet{Geha12} 
find that quenched dwarf galaxies are not identified at separations larger than 1.5 Mpc from an L* galaxy (M$_{\rm star} = 10^{11}$ M$_\odot$). This corresponds to $\Theta \sim -11.5$.

\end{enumerate}
 
This isolation criteria is applied to both the observed and mock galaxy catalogs,
as described below.  The projected relative velocities ($\Delta V_{\rm LOS}$) between each dwarf and each Massive Galaxy are computed as: 

\begin{equation}
\label{eq:delv}
\Delta V_{\rm LOS} = c \frac{ | z_{d} - z_{m} | }{ 1+z_{\rm avg}}  
\end{equation}

\noindent where $c$ is the speed of light, $z_{d}$ is the redshift of the dwarf, $z_{m}$ is the 
redshift of the Massive Galaxy and $z_{\rm avg} = (z_d + z_m)/2.0$.

Projected separations (r$_p$) are determined from the 
angular separation ($A_{\rm sep}$; in radians) between each dwarf and each Massive Galaxy and 
the angular diameter distance $D_A$:

\begin{equation}
{\rm r}_p = A_{\rm sep} \times D_A  \,\, {\rm kpc.} 
\end{equation}

\begin{figure*}
\begin{center}
\mbox{\includegraphics[width=3.35in, trim={0.3in 0in 4.6in 0in}, clip]{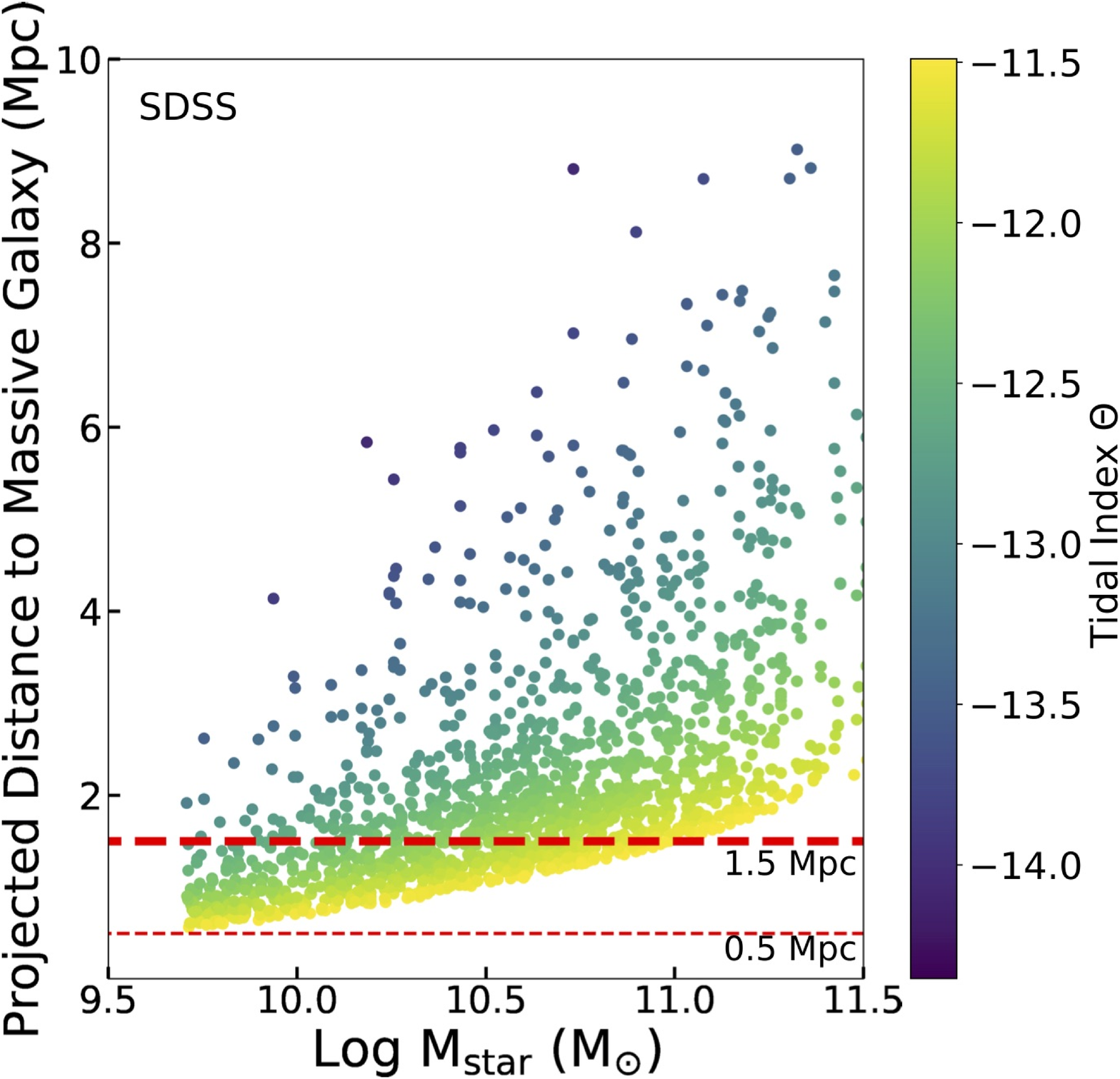}
\includegraphics[width=4in, trim={1.8in 0in 0in 0in}, clip]{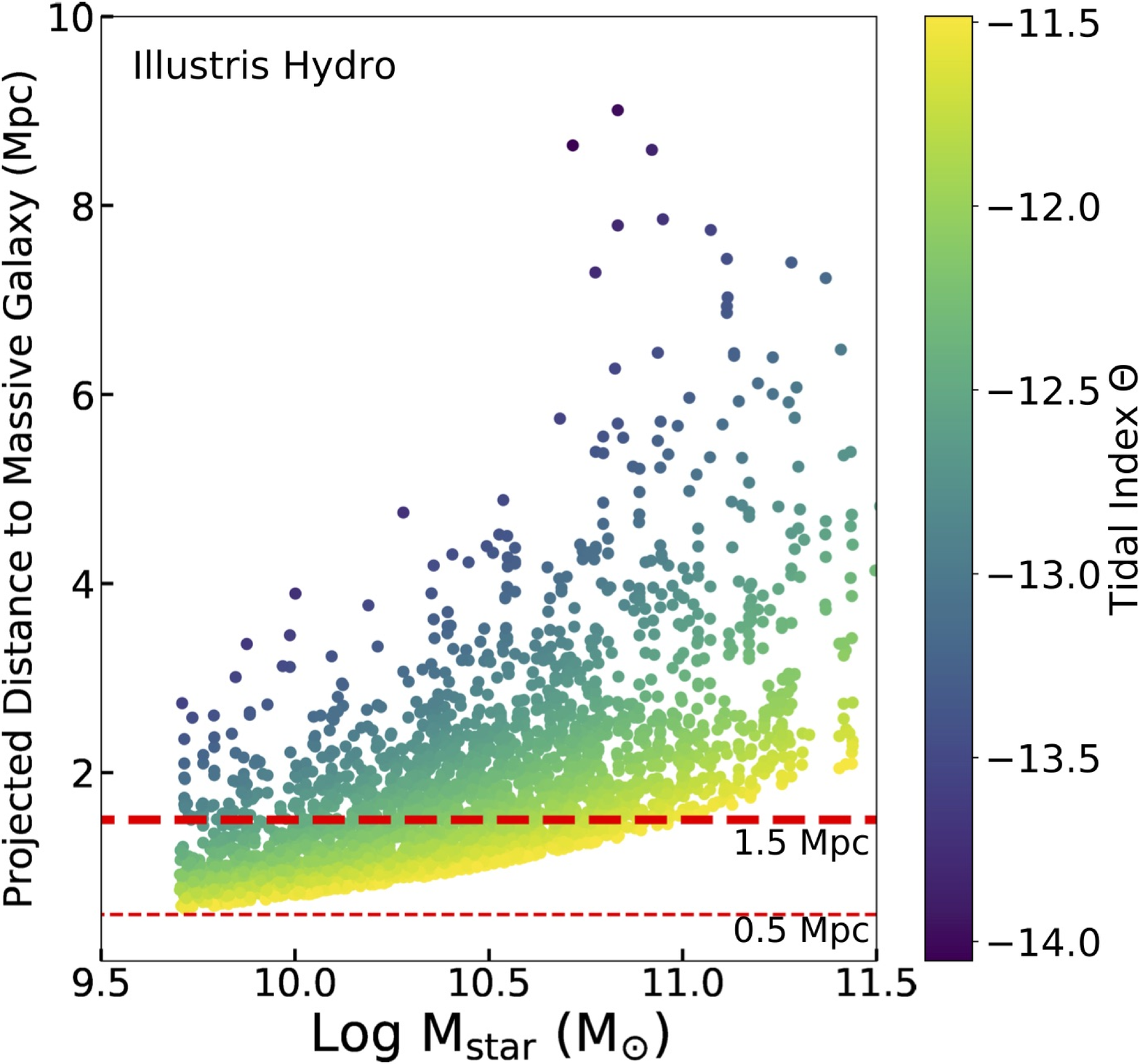}
}
 \end{center}
 \caption{  \label{fig:IsolTidalIndex} 
The stellar mass and separation of the Massive Galaxy (M$_{\rm star} > 5\times 10^{9}$ M$_\odot$) with the largest Tidal Index ($\Theta$, color bar, equation~\ref{eqn:Tidal}) to each {\bf isolated} dwarf identified in the {\it SDSS} (left) and {\it Illustris Hydro} (right) catalogs. Each Massive Galaxy is also required to have a relative velocity of $\Delta V_{\rm LOS} < $ 1000 km/s from each dwarf. All isolated dwarfs are found to be located more than 0.5 Mpc from all Massive Galaxies and more than 1.5 Mpc from all galaxies with M$_{\rm star} > 10^{11}$ M$_\odot$. 
  } 
 \end{figure*}

To ensure that 
 dwarf galaxies located at the {\it SDSS} survey boundaries are properly tested for isolation, the observational ``Massive Galaxy'' catalog is supplemented with galaxies from the NSA catalog that reside outside
the {\it SDSS} continuous survey footprint.  A dwarf galaxy is considered non-isolated and removed from the sample if at least one Massive Galaxy satisfies the above criteria.
For each isolated dwarf we record the Massive Galaxy with the largest Tidal Index that also satisfies $\Delta V_{\rm LOS} < 1000$ km/s.  The stellar mass of these closest hosts and their projected separation to each isolated dwarf is plotted in Fig.~\ref{fig:IsolTidalIndex} (SDSS results are on the left).

For the mock galaxy sample, the Isolation Criteria are applied as follows:

\begin{enumerate}
\item Mock Massive Galaxies are identified within a physical, cubic volume of (20 Mpc)$^3$ centered 
on each mock dwarf. 
\item If this volume extends beyond the boundaries of the simulation, the periodic nature
of the simulation is exploited to create a new (20 Mpc)$^3$ volume centered on the mock dwarf to ensure that 
dwarfs at the simulation boundaries can be properly tested for isolation.
\item The 3D position vector of each mock dwarf and Massive Galaxy from the observer location is translated into 
spherical coordinates, assuming the observer is located at coordinate (0,0,0) in a Cartesian system. 
\item The angular separation between the mock dwarf and each mock Massive 
Galaxy in the (20 Mpc)$^3$ volume is computed and used to determine the projected separation as 
described above. This allows us to compute the Tidal Index $\Theta$.
\item Relative velocities are computed using mock redshifts that account for the peculiar 
motion of the mock galaxies along the line of sight (eqn.~\ref{eq:delv}).
\item If any mock Massive Galaxy is found to satisfy both criteria, the mock dwarf is removed from the sample.
\item If no mock Massive Galaxies are found to satisfy both criteria, then the mock dwarf stays in the sample and the 
separation from the mock Massive Galaxy with the largest Tidal Index and $\Delta V_{\rm LOS} < 1000$ km/s is recorded to double check that isolation is satisfied (see right panel of Fig.~\ref{fig:IsolTidalIndex}). 
\end{enumerate} 

This set of Isolation Criteria ensures that all dwarfs are more than 1.5 Mpc from an L* type galaxy (Fig.~\ref{fig:IsolTidalIndex}; thick red dashed line). 
In practice, these criteria also ensure that none of our dwarfs have a Massive Galaxy within 500 kpc ( Fig.~\ref{fig:IsolTidalIndex}; thin red dashed line). 

\subsection{Redshift Limits and Survey Volume}
\label{sec:Volume}

We define the volume of our survey using redshift limits of 0.013$<$z$<$0.0252. 
The survey volume is limited by: 1) the size of the 
{\it Illustris} volume of (106.5 Mpc)$^3$, which corresponds to a maximal redshift of z=0.0252; 2) efforts to circumvent bias from cosmic variance; and 3)
 the sensitivity limits of {\it SDSS}, as explained below.

\subsubsection{\emph{SDSS} Sensitivity Limits \& Catalog Volume}

The sensitivity limits of the {\it SDSS} catalog are outlined in M14.  The catalog is not complete for
dwarf galaxies in our redshift range (or for larger redshifts than considered here). The observability of dwarfs thus varies with redshift as a function of color 
and stellar mass.

\begin{figure*}
\mbox{\includegraphics[width=3.5in, trim={0.1in 0in 1in 0in}, clip]{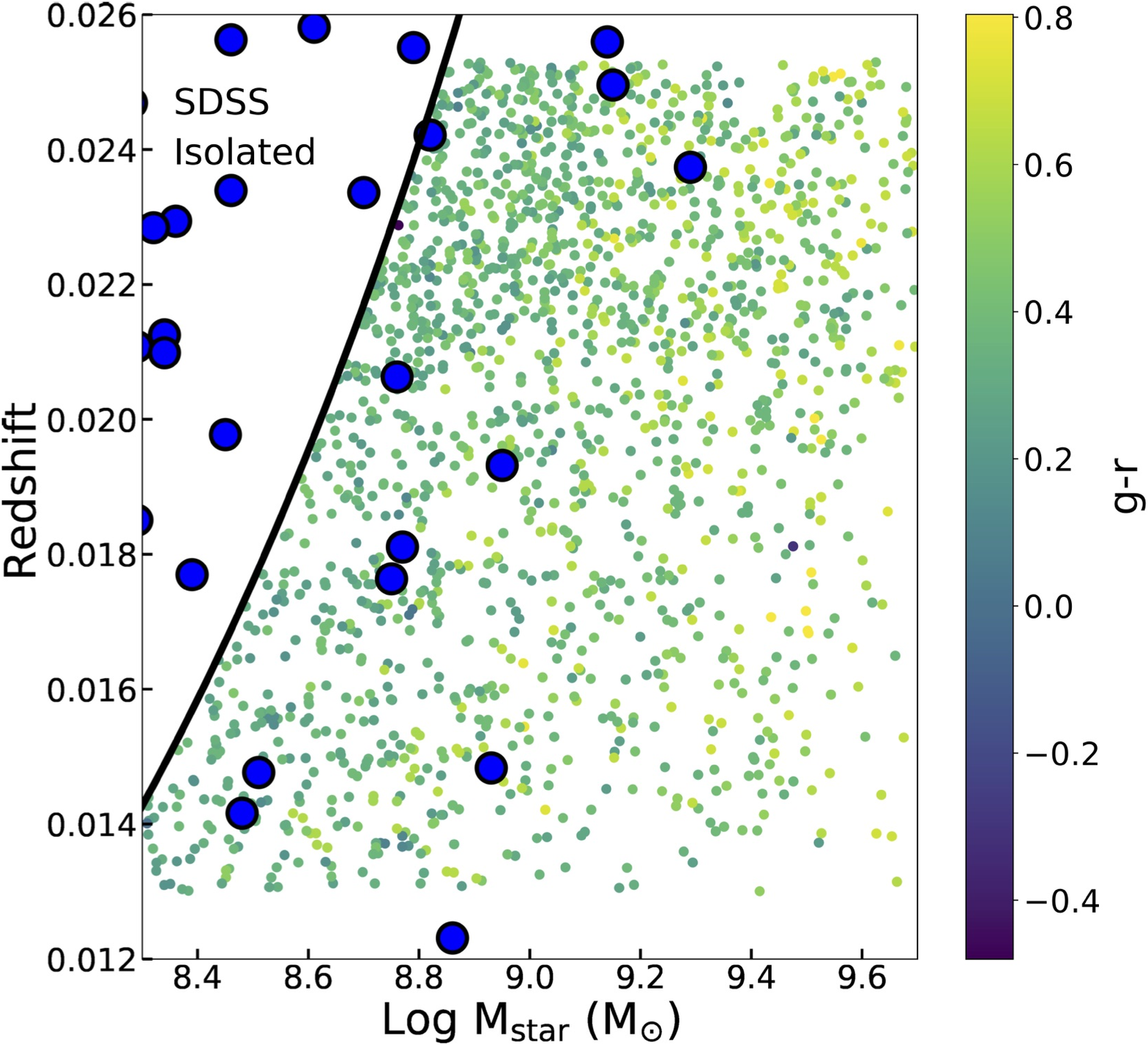} 
\includegraphics[width=3.5in, trim={0.1in 0in 1in 0in},clip]{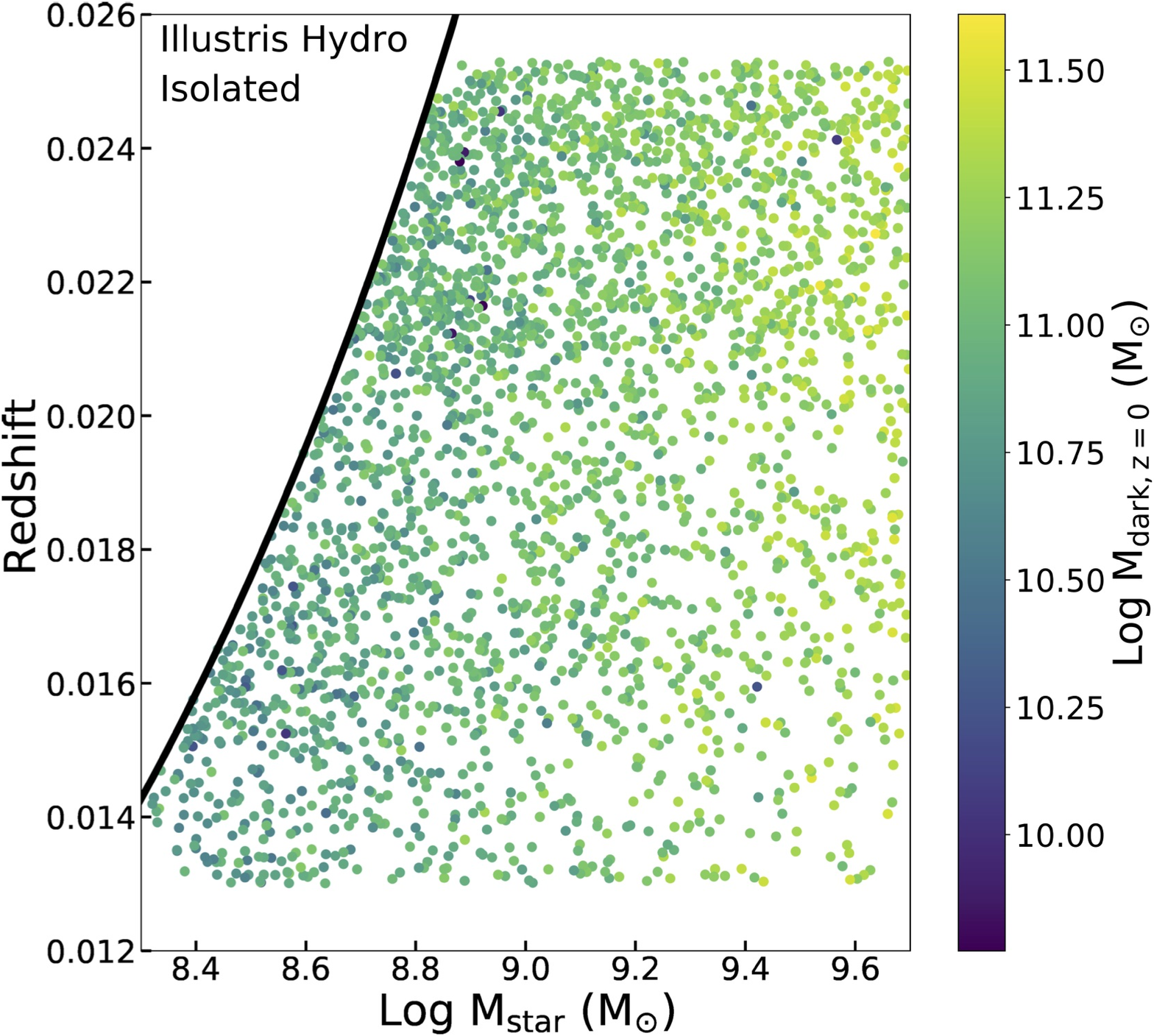}}

 \caption{\label{fig:Mendel} Redshift vs. stellar mass of isolated {\it SDSS} dwarfs ({\it Left}) and mock dwarfs ({\it Right}) that satisfy the redshift and stellar mass limits adopted in this study.
The black solid curve illustrates the adopted sensitivity limits as a function of stellar mass (equation~\ref{eqn:Mendel}).
 All dwarfs plotted here are isolated from more massive galaxies, as defined by the isolation criteria outlined in Section~\ref{sec:Isol}. 
{\it Left:}  Dwarfs are 
 color-coded by restframe $g-r$ color. Blue galaxies are defined as having: $g-r < 0.4$,  Green: $0.4 < g-r < 0.6$ and Red:  $g-r > 0.6$. On average, the {\it SDSS} dwarfs have {\it g-r} $\sim$ 0.41. 
Only 10 of the 28 {\it TNT} dwarfs (larger blue circles)
  would be included in our catalog, and only 2 pairs out of 14 (the primaries 
 are typically selected, but the secondaries fall below the sensitivity limit). 
 {\it Right:}  Results are plotted for one realization of the
  {\it Illustris Hydro} catalog. 
 Points are color coded by the descendent (z=0) halo mass of the mock dwarf. Note that more isolated dwarfs are identified in the mock catalogs than in {\it SDSS}; see Table~\ref{table:IsolCounts}.
  } 
 \end{figure*}

M14 define a fitting function that describes the minimum stellar mass observable for galaxies of a given 
color at a given redshift:  
\begin{equation}
\label{eqn:Mendel}
{\rm log}(M_{lim}/M_\odot)  = \alpha + \beta {\rm log}(z) + \gamma {\rm log}( 1 + z)\,\,\,\ .
\end{equation} 

\noindent We adopt fitting parameters ($\alpha$,$\beta$,$\gamma$):  
\begin{equation}
\alpha = 12.36 \,;
\qquad
\beta = 2.2\,;
\qquad
\gamma = 0.0 \,\,\,\, .
\end{equation}

These parameters correspond to the stellar mass completeness limit for all galaxies bluer than the green valley, as defined by M14. Note that because our sample is low redshift, we chose
$\gamma$=0 \citep[vs. $\gamma = 0.3932$ in][]{Mendel14} neglecting any redshift dependence in the fitting function. 

 These parameters best encompass our desired dwarf sample, which are expected to be blue or green in color 
 as galaxies with stellar masses between $10^8 - 10^9$ M$_\odot$ are found to be typically 
 gas rich and star forming in the field
\citep{Geha12, Bradford15}. 
Furthermore, the average restframe $g-r$ color for the isolated
{\it TNT} dwarf pairs in \citet{Stierwalt15} 
is 0.267, 
with a maximum value of 0.557. In other words, all of the dwarf pairs found in {\it TNT} are either blue or green, but none are red. 

In Fig.~\ref{fig:Mendel} (left) all dwarf galaxies in the M14 catalog that satisfy our redshift and sensitivity limits are plotted as a function of mass and redshift.  Points are color-coded by the dwarf galaxy's restframe $g-r$ value.  
The black solid line illustrates the fitting function from Eqn.~\ref{eqn:Mendel}.
The sample drops off rapidly with redshift towards the low mass end, but blue galaxies are 
observable over a larger mass range.  The resulting average $g-r$ color for the sample of isolated {\it SDSS} dwarfs we consider in this study is $\sim$0.41.

There are 28 dwarfs (14 pairs) in the Isolated {\it TNT} sample that are in the redshift range of our survey (blue circles in the left panel of Fig.~\ref{fig:Mendel}).   
However, given the adopted 
sensitivity limits, roughly 10 of the Isolated {\it TNT} sample (and only 2 pairs) in this redshift range overlap with our 
catalog.
 Many {\it TNT} pairs are missed because the secondary falls below the observability limit. This means that by applying the listed 
 sensitivity floor we can only get a lower limit on the fraction of dwarf multiples.  In Section~\ref{sec:LSST}, we remove these
sensitivity limits but keep a mass floor of $2\times 10^8$ M$_\odot$, which would recover 18 out of the 28 
isolated {\it TNT} dwarfs (but only 8 out of 14 complete pairs). 

The fact that the {\it TNT} pairs have members at masses below our sensitivity limits (black line in Fig.~\ref{fig:Mendel}) strongly indicates that the dwarf multiple fraction determined in this study will significantly underestimate the true dwarf multiple fraction if the 
definition of ``dwarf'' were extended to lower masses.
Our goal in this study is to compute the dwarf multiple fraction using parameters that can be reasonably reproduced by both observations and theory.

\subsubsection{Illustris Simulation \& Mock Catalog Volume}
\label{sec:Mock}

To create the mock catalog, sight lines are drawn through the {\it Illustris} simulation volume starting 
from a random location in the full (106.5 Mpc)$^3$ simulation volume and using a random viewing perspective. 
Because the simulation boundary conditions are periodic, 
the full volume can be recreated regardless of the chosen starting location or viewing orientation. 
Note we could exploit the periodic boundary conditions to 
 expand the volume to larger redshifts \citep[e.g.][]{Snyder17}. However, the sensitivity 
limits of {\it SDSS} at larger redshift (Fig.~\ref{fig:Mendel}) imply that the dwarf statistics would not increase appreciably. 

\begin{figure*}
\begin{center}
\mbox{\includegraphics[width=3.5in, trim={3.5in 3.2in 3.2in 2.5in}, clip]{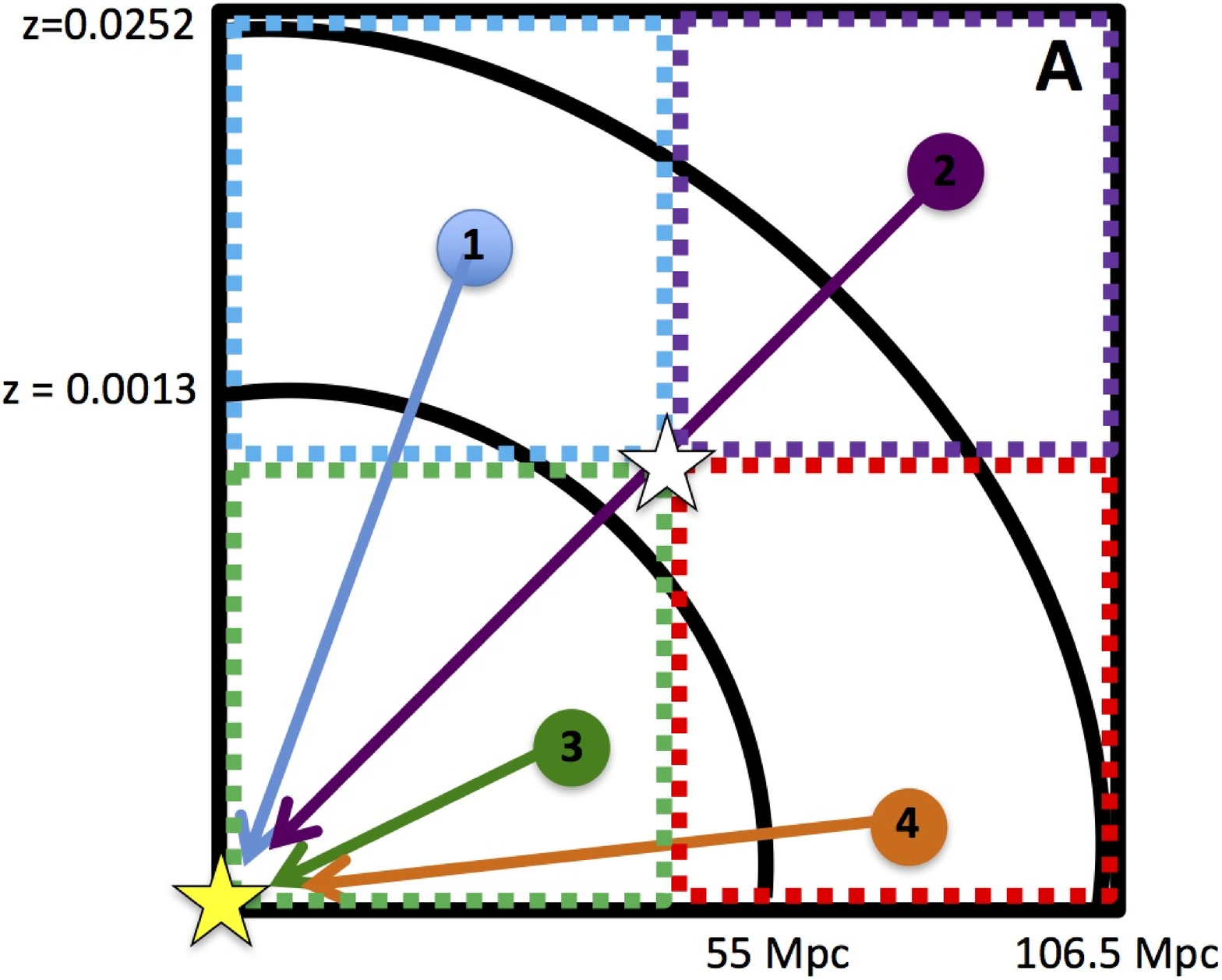}
\includegraphics[width=3.5in, trim={3.5in 3.2in 3.2in 2.5in}, clip]{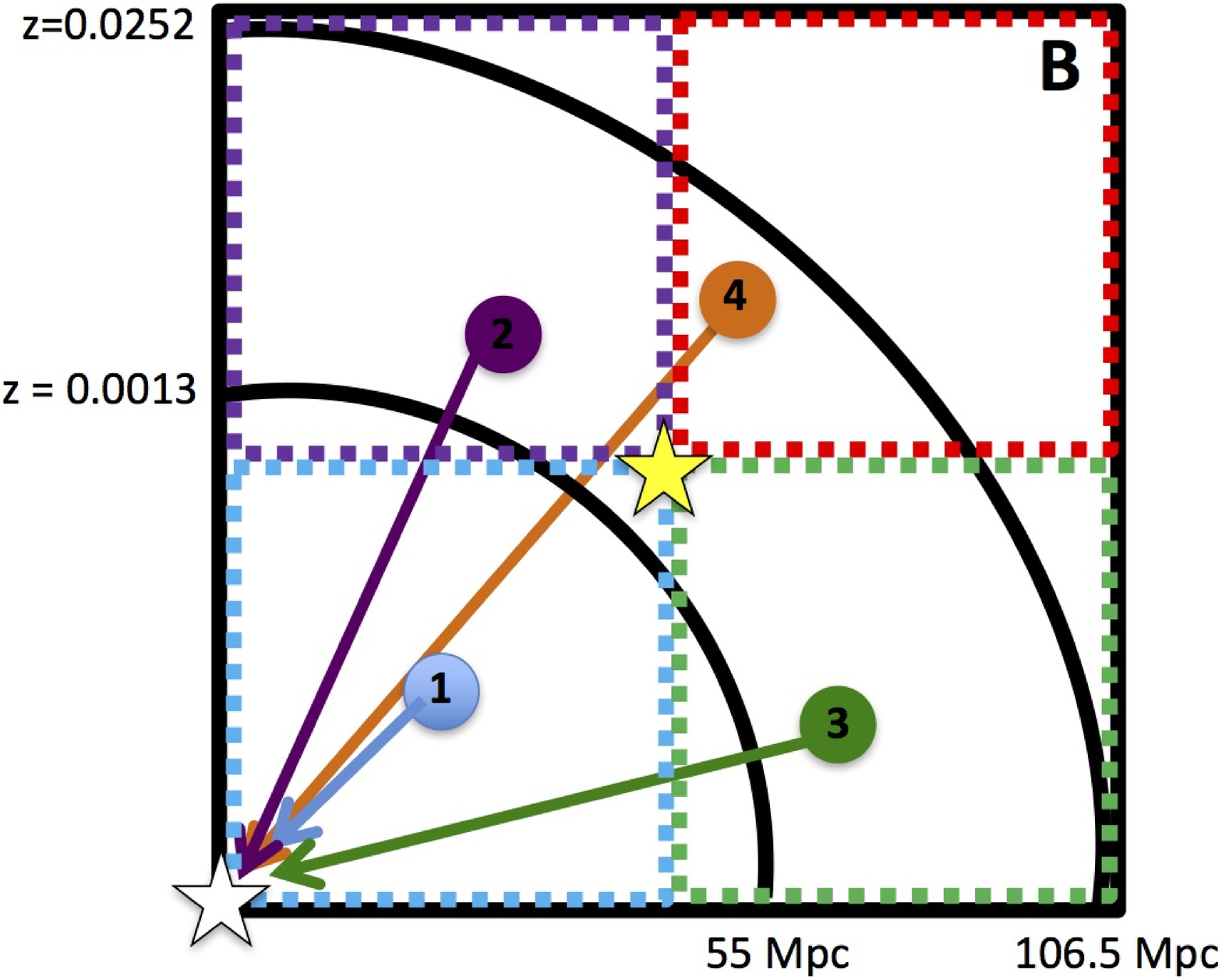}}
 \end{center}
 \caption{\label{fig:Observer}  Schematic illustrating how the simulation volume is sampled to 
 generate `mock' galaxy catalogs. 
 {\bf Schematic A (left)} illustrates a 2D view of the simulation 
 volume for one viewing perspective, where the observer is placed at the bottom left (coordinate 0,0; yellow star).
 Four example subhalos are marked by colored circles, with line of sight vectors drawn to the observer.
 Black arcs indicate the redshift limits defined for the catalog (0.013 $< $z $< $0.0252). 
 In Schematic A, only subhaloes 1 and 4 (blue and orange) would be included in this realization of the mock survey volume.
 {\bf Schematic B (right)}, represents a different realization of the same volume, where both the observer location and viewing perspective have been changed relative to Schematic A. To generate Schematic B, the observer is moved to the center of Schematic A (white star) and we choose a different 
 viewing perspective (the +y axis in Schematic B is the -x axis in Schematic A). The volume is then recreated, with the white star now at coordinate (0,0). Colored dashed boxes are included to mark how the volume has been re-ordered from Schematic A to B.
Given that the simulation boundary conditions are periodic, the volume can be reshuffled self-consistently - in the 
new realization of the volume, the observer in A (yellow star, at 0,0) is now at the center of volume B.
The survey redshift limits are redrawn (black arcs) and new redshifts are defined in volume B for 
each subhalo. 
The new 
survey volume in Schematic B thus encompasses different regions of the simulation volume than that of Schematic A. 
In particular, now subhaloes 2, 3 and 4 are included in the catalog (purple, orange and green).  
This process is repeated 500 times to generate the statistics quoted in this study. 
  } 
 \end{figure*}

As indicated in Fig.~\ref{fig:Observer}, 
an `observer' is placed at a randomly-selected starting location with a randomly-chosen viewing direction. 
The 3D distance is computed between the observer and each subhalo as well as the velocity of each subhalo along the line-of-sight to the observer ($v_{\rm los}$).

Only those subhaloes with redshifts within the 
specified redshift limits are included in the analysis. 
This process is repeated 500 times, adopting a new observer location 
and viewing perspective each time (e.g. Schematic B in Fig.~\ref{fig:Observer}). All statistics presented in this study for the `mock' galaxy 
catalogues are averaged over the 500 different realizations of the survey volume.   

We have adopted a lower redshift limit of z$>$0.013, which corresponds to a comoving distance of $\sim$55 Mpc.  
This was chosen so that only $\sim$half of the entire simulation volume is utilized for the selection of galaxies for a given 
combination of observer location and viewing perspective. As a result, we can sample a different volume each time
the mock galaxy catalog is generated.  
This allows us to address biases in our statistics introduced by 
 cosmic variance and sampling bias. We are sampling different survey volumes that are 
affected by different large scale structures, rather than sampling the same volume repeatedly.

With redshift limits of 0.013 $< z < $0.0252 the corresponding volume for our observational sample in {\it SDSS} 
is $\sim$(95 Mpc)$^3$. For the {\it Illustris} simulations, this redshift range corresponds to a volume of  
$\sim$ (81 Mpc)$^3$, a factor of $\sim$1.4 times smaller than the observational volume and a factor of $\sim$ 2 smaller 
than the total simulation volume.  In other words, the adopted mock survey volume covers 1/8th of the 
sky (0.5$\pi$ steradians), while {\it SDSS} covered $\sim$18\% of the sky, which yields a ratio of $\sim$1.4.  
This survey volume is significantly smaller than that sampled by the {\it TNT} survey \citep[0.005$<$ z $<$ 0.07;][]{Stierwalt15}, 
but affords us the fairest comparison between the {\it SDSS} catalog and the {\it Illustris} simulation volume.

The mock catalog generation includes peculiar motions when determining the redshift and is thus consistent with the measurement of observational redshifts, as it is the sum of the Hubble flow and the peculiar velocity.
Note that peculiar motions may cause us to assign companions to dwarfs that are actually at much larger 3D separations. This should affect both the mock and observed galaxy catalogs (see Section~\ref{sec:DwarfCompanion}).  

At this point, the mock galaxy sample is complete at all stellar masses within the mock survey volume. 
However, as described in the previous section, {\it SDSS} sensitivity limits imply a strongly increasing 
incompleteness as a function of increasing redshift and decreasing stellar mass (Fig.~\ref{fig:Mendel}; left). 
To ensure the mock and observed dwarf galaxy catalogs are comparable, we apply the same 
sensitivity limits, as a function of redshift and stellar mass,
 to both the observed and mock galaxy catalogs, as described by Eqn.~\ref{eqn:Mendel}. 
 As a result, the mock catalogs will be similarly incomplete as a function of stellar mass (see right panel 
 of Fig.~\ref{fig:Mendel}).

\section{Results: Galaxy Counts within the Observational and Mock Survey Volumes} 
\label{sec:NCount}

\subsection{Total Galaxy Counts: Massive Galaxies \& Dwarfs}
\label{sec:NCountAll}

Baryonic effects (stellar feedback, reionization) 
can dramatically affect galaxy counts at the low mass end 
\citep{Chua17,Sawala13, Cui12,Kravtsov10}.  
Here we compare galaxy counts within our survey volume (0.013 $< z < $0.0252)
for massive galaxies (M$_{\rm star}> 5 \times 10^9$ M$_\odot$)
 to ensure the observational
and `mock' galaxy catalogs yield consistent results in a mass regime where baryonic effects 
should not play a significant role  
\citep[i.e. a  mass regime where the stellar-halo mass relations are well calibrated][]{Genel14}. 
We then extend this analysis to the dwarf galaxy
regime.  

For the `mock' galaxy catalogs we quote statistics averaged over 500 sightlines that randomly sample
the {\it Illustris Hydro}
simulation volume, as described in the previous section and by 
Fig.~\ref{fig:Observer}.
For the {\it SDSS} dwarf galaxy counts, we have randomly sampled the stellar mass
and redshift error space  
assuming Gaussian errors from M14 and {\it SDSS} DR9, respectively, to generate 
500 versions of the dwarf galaxy catalog. 
 Galaxy counts are computed as the mean over these catalog realizations and the quoted errors are 
 the standard deviations. Results are summarized in Table~\ref{table:Counts}.
 
 Note that we have not generated multiple versions of the massive galaxy 
catalog (M$_{\rm star} > 5 \times 10^9$ M$_\odot$) as the {\it SDSS} DR9 survey is
 complete with respect to stellar mass for such galaxies over the redshift range considered. 
 Given the small standard  
  deviation in dwarf galaxy counts, this omission is not expected 
 to significantly affect the galaxy counts of massive galaxies listed here. 
  Also, as described in Section~\ref{sec:ObsSample}, we supplement the observed catalog with 
 massive galaxies identified in the NSA catalog.

\begin{table*}
\centering
\caption{ {\bf Dwarf and Massive Galaxies in the \emph{SDSS} and mock survey volumes.} 
\newline  All values are the mean number counts and standard deviations computed over 500 realizations of the respective catalogs. 
{\it Columns 2 and 4)} {\it SDSS} values account for a spectroscopic completeness of 88\%. Note that the stellar masses for the {\it SDSS} massive galaxies are not randomly sampled, and so the number count and density are listed without uncertainties.   
{\it Columns 3 and 5)} $n_{\rm massive}$ and $n_{\rm dwarf}$ refer to the number density of Massive and dwarf galaxies, respectively. }
\label{table:Counts}
\begin{tabular}{ |c|c|c|c|c|c| }
\hline\hline
Catalog &  \# Massive Galaxies & $n_{\rm massive}$ (Mpc)$^{-3}$  & \# Dwarf Galaxies & $n_{\rm dwarf}$ (Mpc)$^{-3}$ & Volume (Mpc)$^{3}$ \\
\hline
SDSS &   6,944 &  0.010 &  7860 $\pm$ 35  &  0.010 & 7.78 $\times 10^5$  \\
Illustris Hydro  &  4,810 $\pm$ 741 &  0.009 $\pm$0.001 & 9,165 $\pm$ 1,607 & 0.017$\pm$0.003 &  $5.50 \times 10^5$ \\
\hline\hline
\end{tabular}
\end{table*}

From Table~\ref{table:Counts}, the number densities for Massive Galaxies are consistent between 
observations and theory, as expected as the feedback prescriptions adopted in {\it Illustris} are 
calibrated 
such that the stellar-halo mass relation agrees with 
the observed number counts of Massive Galaxies \citep{Vogelsberger14}.  The difference in the total number 
counts reflects the factor of 1.4 larger volume of the {\it SDSS} footprint vs. the {\it Illustris} volume.  

The dwarf number densities are more discrepant, but still agree within 2$\sigma$. On average there 
appear to be more mock dwarfs than in {\it SDSS}, which is likely a manifestation of 
 the missing satellite problem. 
  Many of these dwarf subhaloes are in proximity to a 
 massive host, and would likely be quenched by environmental processes, and therefore
  unobservable at low masses given our optimistic cuts in color for the observability of dwarfs in {\it SDSS} 
  (see Fig.~\ref{fig:Mendel}). 
In the next section we apply an isolation criterion to identify dwarfs that would still be star forming and thus bluer in color
in order to provide a fair comparison between the observations and the simulation data.

\subsection{Isolated Dwarf Galaxy Counts} \label{sec:PropIsol}

In this section the properties (number counts, mass distributions) of mock and observed 
dwarfs that are isolated from Massive Galaxies are quantified and compared for consistency.
Recall, the distribution of isolated {\it SDSS} dwarfs are plotted as a function of redshift in the left panel of Fig.~\ref{fig:Mendel}, with mock catalog results on the right.

Properties of the isolated dwarfs are summarized in 
Table~\ref{table:IsolCounts}. As expected, the number density of isolated dwarfs is much lower than that of non-isolated dwarfs
in the observed and both mock galaxy samples  (factor of 3 and 2, respectively). However, 
 the mean number density of dwarfs in the mock catalog is still a factor of 2 larger than that observed in {\it SDSS}. Note that the results do still agree within 2$\sigma$ and the minimum density of mock isolated 
dwarfs identified across all 500 realizations of the {\it Illustris} volume is 0.003 (Mpc)$^{-3}$, in agreement with the observed value. 
Despite the higher number density of dwarfs in the mock catalogs the mean stellar mass of isolated dwarfs is consistent with observations; all catalogs yield $\big <\log$(M$_{\rm star}/$M$_\odot) \big >$ = $9.1 \pm 0.3$.

\begin{table}
\centering
\caption{ {\bf Number of dwarf galaxies in the observed and mock survey volume that are isolated from all massive galaxies (M$_{\rm star} > 5 \times 10^9$ M$_\odot$).} 
\newline All values are the mean and standard deviation computed using 500 realizations of the respective catalog.
 {\it Columns 2 and 3)}  {\it SDSS} DR9 observed counts are listed accounting for a spectroscopic completeness of 88\%.
{\it Column 3)} $n_{\rm dwarf iso}$ refers to the number density of isolated dwarfs in the survey volume, following criteria listed in Section~\ref{sec:Isol}.}
\label{table:IsolCounts}
\begin{tabular}{ |c|c|c| }
\hline\hline
Catalog & \# Isolated Dwarfs & $n_{\rm dwarf iso}$ (Mpc)$^{-3}$  \\ 
\hline
SDSS &   1,909 $\pm$ 15  &  0.00245 $\pm$ 0.00002 \\ 
Illustris Hydro & 2,829 $\pm$  422  &  0.0051 $\pm$ 0.0008
\\ 
\hline\hline
\end{tabular}

\end{table}

On average, there is a factor of $\sim$2 more mock dwarfs
than observed in {\it SDSS} in a given volume (Table~\ref{table:IsolCounts}). 
 There are a number of possible explanations for this discrepancy, e.g.: 
\begin{enumerate}
\item  There are more low surface brightness dwarf galaxies for the same stellar mass that are not currently observable given 
 the sensitivity of {\it SDSS} \citep{Klypin15, Blanton05}.
 Their 
 detection might be possible with new observing strategies \citep{Greco17, Danieli18} or when surveys like {\it LSST} come online. 
\item Environmental effects
 are not the only explanation for the subhalo overabundance problem. \citet{Wetzel16} have shown that different choices in star formation and feedback prescriptions can help
remedy the problem. We do not explicitly account for the stellar mass formed in the simulation and thus do not account for this effect. However, while we are only exploring a factor of 10 in mass, it is not clear that such processes would similarly affect mass growth across this dwarf mass regime \citep{diCintio14}. 
\item Reionization may hinder early stellar mass growth, reducing the number of observable low mass galaxies across the mass range probed \citep[e.g.][]{Wyithe06}. However, it is unclear how well this would work at LMC mass ranges. 
\item There are fewer massive hosts at cluster mass scales (M$_{\rm star} \sim10^{12}$ M$_\odot$) 
 in the {\it Illustris} volume than in our selected {\it SDSS} volume (which is 1.4 times larger). In fact there are no systems at
  this stellar mass
 scale in our mock Massive Galaxy catalogs, but there are 3 in our {\it SDSS} Massive Galaxy Catalog.  The Isolation Criteria 
 will thus remove more dwarfs in the observed catalog than in the mock catalog. 
 \end{enumerate}
 
 These 4 possibilities may be testable in next generation versions of cosmological simulations - particularly ones with larger 
 volumes and different baryonic physics \citep[e.g. Illustris-TNG,][]{Springel17}. {\bf It remains a challenge to explain how these processes would uniformly affect the number density across the mass range probed in this study.}

\begin{figure}
\begin{center}
\mbox{\includegraphics[width=3.5in]{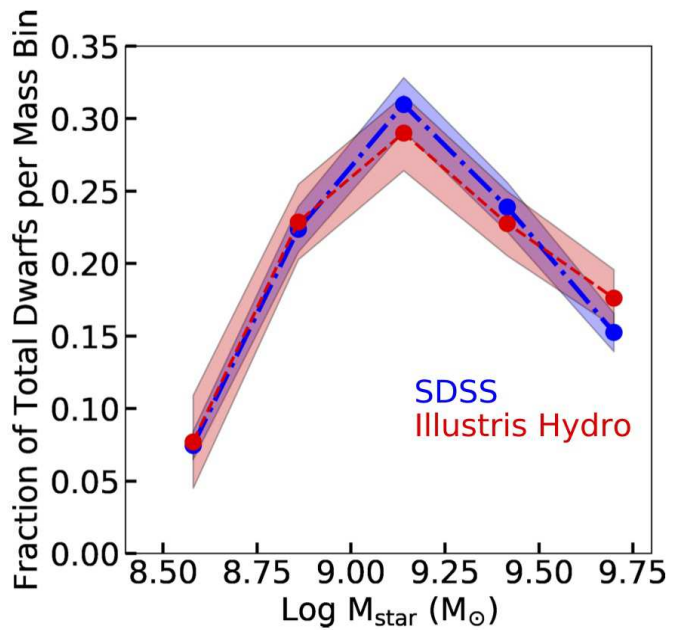}}
 \end{center}
 \caption{\label{fig:GalMass} 
 Number of isolated dwarf galaxies per stellar mass bin, normalized by the total number of 
 dwarf galaxies 
 in the entire respective catalog. 5 mass bins are equally spaced in $\log$(M$_{\rm star}$) intervals of 0.28 starting at 
 8.3 to 9.7. Circles mark the end
 of each mass bin and the mean value of the ratio. Shaded regions indicate 2$\sigma$ errors. 
 The {\it SDSS} results are denoted by the blue dash dotted line. 
 {\it Illustris Hydro} results are plotted in red (dashed), showing good agreement.  The rapid decline in all catalogs at low masses
is a result of the sensitivity limits (Eqn.~\ref{eqn:Mendel}).
  } 
 \end{figure}

In Fig.~\ref{fig:GalMass} the fraction of mock and observed isolated dwarf galaxies is plotted 
for 5 equally spaced mass bins.
Thus, although the total number of dwarf galaxies in the mock catalogs is higher than observed, 
the fraction of dwarfs per mass bin 
shows very good agreement.   Since we are concerned with the ratio of dwarfs with companions
relative to the total dwarf population, we expect that such ratios will be physically meaningful.

If the solution to the discrepancy between theory and observations is missing low surface brightness galaxies, Fig.~\ref{fig:GalMass}  suggests, perhaps surprisingly, that a similar fraction of low surface brightness 
dwarfs are missing across all mass bins.

\section{Results: Frequency of Dwarf Multiples and Predictions for Next Generation Surveys}
\label{sec:FreqPairs}

In the following sections we quantify the frequency of isolated dwarf multiples, where all galaxy members have stellar
masses from $2 \times 10^8$ M$_\odot <$ M$_{\rm star} < 5 \times 10^9$ M$_\odot$ and none are in proximity to a Massive Galaxy (M$_{\rm star} > 5 \times 10^9$
M$_\odot$).  

In Section~\ref{sec:DwarfCompanion}, we define our selection criteria for dwarf multiples.  In Section~\ref{sec:Nc}, the 
 mean number of companions per dwarf galaxy ($N_c$) is computed for all catalogs and calibrated for projection effects (multiples with 3D separations less than 300 kpc) using the cosmological catalogs.
 This is further quantified as a function of dwarf mass, $N_{c,m}$ in Section~\ref{sec:Ncm}.
 In Section~\ref{Freq:Pairs}, we compute the frequency of dwarf pairs vs. triples or higher order multiples for dwarf galaxies in the adopted mass range. Finally, in Section~\ref{sec:LSST} we calibrate the {\it SDSS} results for completeness using the mock catalogs in order to make predictions 
 for future surveys that will be complete to stellar masses as low as 2 $\times10^8 $ M$_\odot$ out to 100 Mpc.


\subsection{Identification of Isolated Dwarf Multiples} 
\label{sec:DwarfCompanion}

We seek to identify dwarf galaxies with close companions that have 
separations and relative velocities that are plausible for tidally-interacting, bound systems
of this mass scale.
Informed by results from the {\it TNT} surveys and numerical simulations of the 
Magellanic Clouds, 
dwarfs are deemed associated if they have a relative line-of-sight velocity  
less than 150 km/s, angular separation $< 55 \arcsec$  and a projected separation r$_p <$150 kpc. {\it SDSS} fiber collisions reduce the number of detectable companions 
at angular separations $< 55 \arcsec$, which corresponds to projected separations of $\sim$15-25 kpc over the redshift range probed in this study.

In {\it Illustris} the gravitational forces exerted by dark matter particles
are softened on a comoving scale of $\epsilon = 1.4$ kpc.
We are thus unable to resolve subhaloes that are separated by less 
than roughly 5 softening lengths, corresponding to a separation of 
roughly 11 kpc at z=0.  In practice there are very few dwarf pairs that 
are identified with physical separations less than 15 kpc - this is likely due to SUBFIND's
 inability to distinguish haloes separated by such small distances \citep[see, e.g.][]{Rodriguez15}.
Since the projected separation is always smaller than the 3D separation, by applying a 
limit of 55$\arcsec$ on the angular separation, we ensure 3D separations large enough to avoid this issue.

The projected separation upper limit is roughly $R_{\rm 200}$ 
for the most massive dwarf subhaloes in the mock catalog ($M_{\rm dark} \sim M_{\rm 200} \sim 
4.0 \times 10^{11}$ M$_\odot$; see Fig.~\ref{fig:DMStars}).  Furthermore, \citet{Stierwalt15} found 
 that SFRs in dwarf galaxy pairs are elevated relative to non-paired 
 dwarfs even at separations of 120 kpc.  An upper limit of 150 kpc will thus ensure that 
 all plausibly interacting dwarfs are identified.

The upper limit on the line of sight velocity corresponds to the escape speed of our 
most massive dwarfs at a distance of 150 kpc.   The selection criteria 
thus ensures that bound dwarf pairs will be captured even at large separations.
For reference, the 3D velocity difference between the Magellanic Clouds is 
$\sim$130 km/s \citep{Kalli13} and their 3D separation is $\sim$23 kpc; our 
criteria would allow for the selection of such analogues.

Note that these criteria differ from those adopted by the {\it TNT} survey \citep{Stierwalt15}, where 
dwarf pairs were selected to have separations less than 50 kpc and relative velocities 
less than 300 km/s (although the majority of the sample have velocities less than 150 km/s).  We adopt a larger separation limit because of the issues with close 
separations outlined above. We also adopt a lower relative velocity limit in order to 
minimize the frequency of chance projections.

\subsubsection{Selection of \emph{Physical} and \emph{Projected} Dwarf Multiples} 
\label{sec:SelectionP}

Using the defined {\it SDSS} and mock, isolated dwarf galaxy catalogs, {\bf ``\emph{Projected}''} dwarf multiples are identified using the 
following steps based on their projected separations and relative line-of-sight velocities.

\begin{enumerate}
\item All isolated dwarf galaxies are rank-ordered by stellar mass.
\item Starting with the most massive dwarf, the projected separation and line-of-sight velocity difference
are computed between that dwarf and every other dwarf in the isolated catalogs.
\item All other dwarf galaxies located with angular separations $< 55\arcsec$, projected separations of r$_p < 150$ kpc, and $\Delta V_{\rm LOS} < 150$ km/s of the given dwarf
are stored as companions. The angular separation limit is applied to all catalogs to avoid incompleteness owing to fiber collisions in {\it SDSS} and create a similarly constrained mock catalog.
\item The above steps are repeated for the next most massive dwarf.  
\end{enumerate}
All steps are repeated for 500 realizations of the {\it SDSS} and {\it Illustris Hydro} 
 catalogs. The resulting catalog of multiples will be referred to as {\bf \emph{``SDSS''}} and {\bf \emph{``Illustris Hydro Projected''}} in subsequent plots.

In Section~\ref{sec:PropIsol}, it was shown that the number density of isolated mock dwarfs is roughly 
a factor of two larger than that observed (Table~\ref{table:IsolCounts}).  
While the relative fraction of isolated dwarfs is consistent between theory and observations 
(Fig.~\ref{fig:GalMass}), it is possible that the higher density of dwarfs will result in a larger frequency 
of dwarf multiples if they are selected based on projected properties alone. 

To address this issue we consider a second method for identifying mock dwarf multiples with small 3D separations. 
Before step 2 in the methodology outlined above, an additional step is introduced.  Starting with the 
most massive dwarf, we first identify all isolated mock dwarf galaxies located within a 3D sphere
 300 kpc in radius centered on the target mock dwarf and then proceed to step 3.  
 This ensures that the contamination fraction from 
 projection effects will be minimized. In contrast, the {\it Illustris Hydro Projected} sample may contain 
 mock dwarfs separated 
 by much larger 3D distances that appear to be in close proximity owing to projection effects.  
 This process is repeated for 500 realizations of the dwarf mock catalog to create a sample that will be 
 referred to as {\bf \emph{``Illustris Hydro Physical''}} in subsequent plots.

\begin{figure*}
\begin{center}
\mbox{\includegraphics[width=2.52in, clip=true, trim=0in 0in 0.98in 0in]{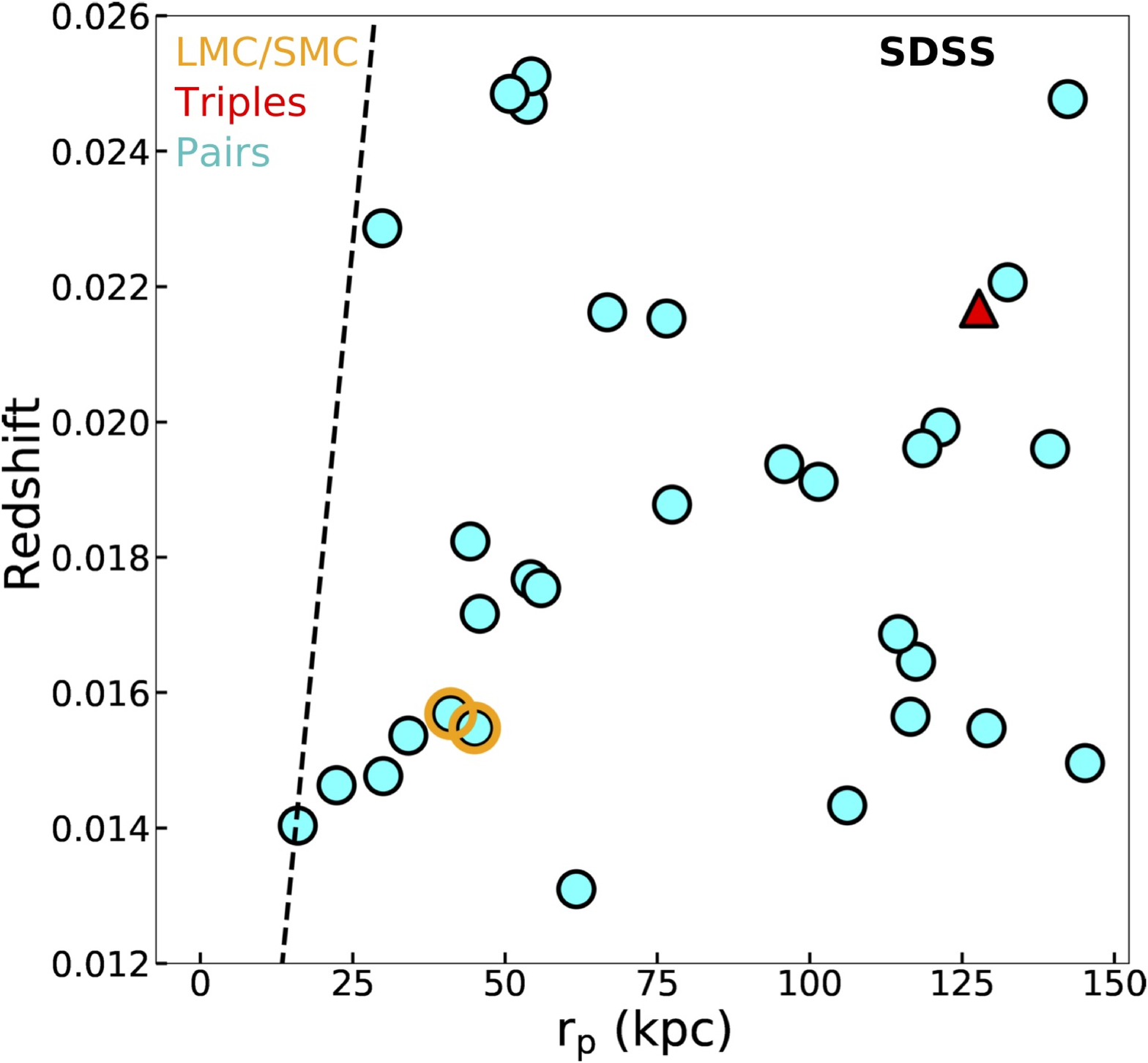}
\includegraphics[width=2.2in, clip=true, trim=2.75in 0in 0.98in 0in]{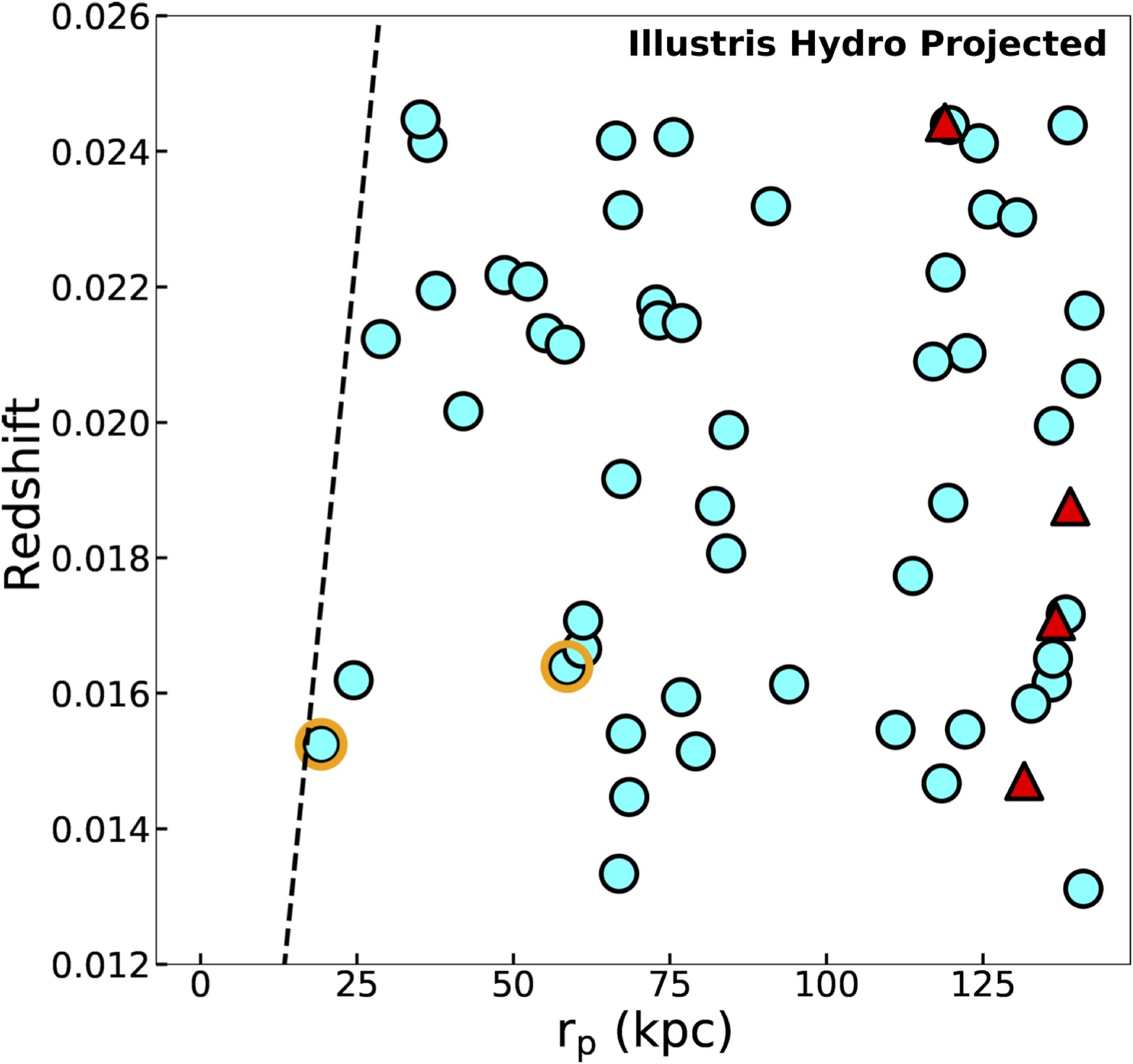}
\includegraphics[width=2.2in, clip=true, trim=2.74in 0in 0.97in 0in]{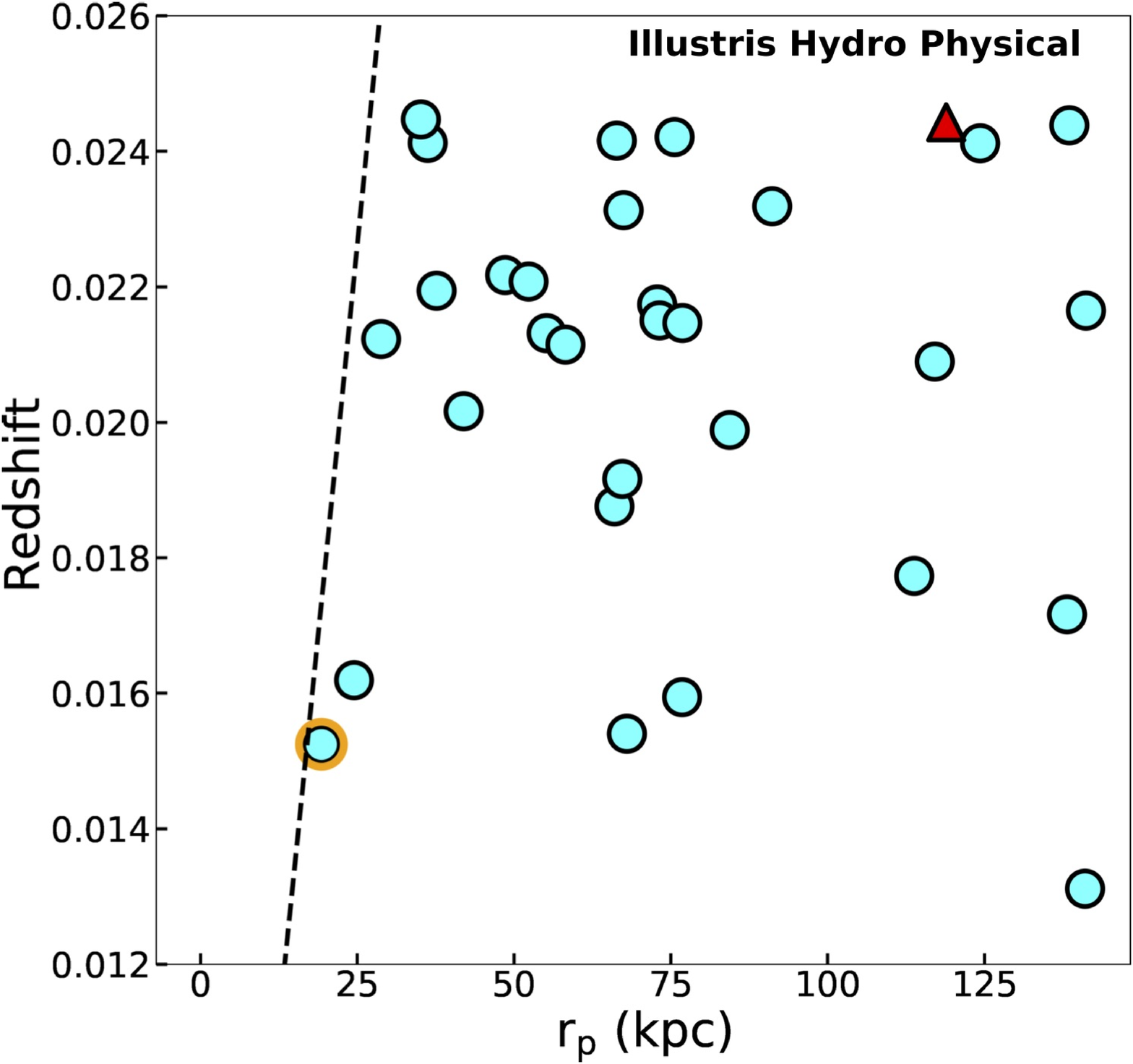} 
}
\end{center}
\caption{\label{fig:RpVSZ} The projected separation between the primary of a dwarf multiple (i.e. most massive dwarf)
and the second (cyan circles) or third (red triangles) member is plotted vs. the average redshift of the primary-secondary or primary-third combination. Unique multiples are plotted for one representative realization of the {\it SDSS} (left) and 
mock catalogs, {\it Illustris Hydro Projected} (center) and {\it Illustris Hydro Physical} (right). 
The dashed black line indicates an angular separation of 55$\arcsec$ at each redshift. All dwarf companions are required to have 
separations from the primary that are larger than this value at a given redshift.
LMC-SMC analogues are marked by orange circles (see Section~\ref{sec:LMC}). Dwarf triples identified at large projected separations ($r_p >$100 kpc) are found to have 3D separations greater than 300 kpc, failing the {\it Physical} criteria (2 red triangles in middle panel, vs. 1 in right panel).  The identification of multiples with small projected separation, like Magellanic Cloud analogues (r$_p < 100$ kpc) and the {\it TNT} sample (r$_p < 50$ kpc), appears robust against projection effects. 
} 
\end{figure*}

Examples of the resulting distribution of dwarf multiples for the three catalogs ({\it SDSS, Illustris Hydro Projected, Illustris Hydro Projected}) are plotted in Fig.~\ref{fig:RpVSZ}.

This figure illustrates the projected separation between each member of the multiple and the primary (most massive) dwarf galaxy member (excluding the primary itself).  Results are plotted for one representative realization of each catalog.  
The solid line indicates the observational limitations in resolving close pairs owing to {\it SDSS} fiber collisions. Since we have 
enforced a minimum angular separations of 55$\arcsec$, no multiples lie to the left of the solid line. When a 3D separation of 
300 kpc is enforced ({\it Physical} criteria; right panel in Fig.~\ref{fig:RpVSZ}), most triples at wide separations are no longer identified. In contrast, multiples with separation less than 50 kpc are found to be robust against projection effects. The {\it TNT} dwarf pairs were all chosen to have 
separations less than 50 kpc; this study indicates that the {\it TNT } dwarf pairs are likely to be true pairs.
 
By repeating this process over multiple realizations of each catalog, we will compute average statistics for the frequency of dwarf multiples, the number of members,  stellar mass 
distributions, and kinematic properties.

\subsection{Mean Number of Companions per Dwarf Galaxy ($N_c$)}
\label{sec:Nc}

Following \citet{Patton00}, we define $N_c$ as the mean number of companions per galaxy. $N_c$ is  
the total number of isolated dwarf companions (i.e. the sum of all pairs, triples, etc., that satisfy the criteria for multiples defined in 
Section~\ref{sec:DwarfCompanion}), divided by the total number of isolated dwarf galaxies in each catalog. 

$N_c$ is computed for all isolated dwarf galaxy multiples in each realization of the {\it SDSS} and mock catalogs ({\it Illustris Hydro Projected} and {\it Illustris Hydro Physical}). The results are then averaged over all 500 realizations of each catalog.
The resulting mean $N_c$ 
and standard deviation for each catalog are summarized in Table~\ref{table:Nc}.  

Agreement is best seen between the {\it SDSS} and {\it  Illustris Hydro Projected} catalogs (agreement is within 1$\sigma$).  
In Section~\ref{sec:Isol} concerns were raised that because the number density of mock dwarfs 
is too high relative to observations, the frequency of mock projected pairs would also be too high. 
However, we have shown here that the fraction of multiples is not overproduced, just as the fraction of dwarfs per mass bin is also in agreement between observations and theory (Fig.~\ref{fig:GalMass}).

When the {\it Physical} selection criteria are applied to the mock catalog, Table~\ref{table:Nc} and Fig.~\ref{fig:Group} illustrate that 
the average $N_c$ decreases by $\sim$40\%. 
In other words, $\sim$40\% of {\it SDSS} multiples may have 3D separations larger than 300 kpc, despite their apparent proximity in projection. 
\citet{McConnachie08} find a similar contamination fraction when identifying massive, compact galaxy groups in the Millennium simulation.
This contamination fraction is also consistent with the study of  \citet{Wilcots04}, who found that a significant fraction of nearby dwarf galaxies with a projected dwarf companion show no strong signs of tidal disturbance in their outer HI structure.

We utilize the theoretical contamination fraction to calibrate the {\it SDSS} $N_c$ for projection effects \citep[see also][]{Patton08}, resulting in the prediction that $N_c$ = 0.024 for the mean number of companions per dwarf with 
physical separations less than 300 kpc (see Table~\ref{table:Nc}).

\begin{table}
\centering
\caption{{\bf $N_c$: Mean Number of Dwarf Companions per Dwarf }
\newline {\it Column 2)} Method for defining multiples (see Section~\ref{sec:SelectionP}). 
{\it Column 3)} The mean $N_c$ and standard deviation computed by averaging over 500 realizations of each catalog.
 {\it The Last Row)} indicates the expected mean fractional number of {\it Physical} companions in the {\it SDSS} sample. The {\it SDSS} $N_c$ is corrected for contaminants owing to projection effects using the fractional difference between the {\it Physical} and {\it Projected} results of the {\it Illustris Hydro} mock galaxy catalog.}
\label{table:Nc}
\begin{tabular}{ |c|c|c| }
\hline\hline
Catalog  & Method & $N_c$  \\
\hline
SDSS &   Projected  &   0.039 $\pm$ 0.003 \\
Illustris Hydro  &  Projected &  0.034 $\pm$ 0.005   \\
Illustris Hydro  &  Physical &  0.021 $\pm$ 0.003    \\
\hline
SDSS-Correction   &  Physical &  $\sim$0.024 \\
\hline\hline
\end{tabular}
\end{table}

\subsection{Mean Number of Companions per Dwarf per Mass Bin ($N_{c,m}$)}
\label{sec:Ncm}

\begin{figure}
\begin{center}
\includegraphics[width=3.5in]{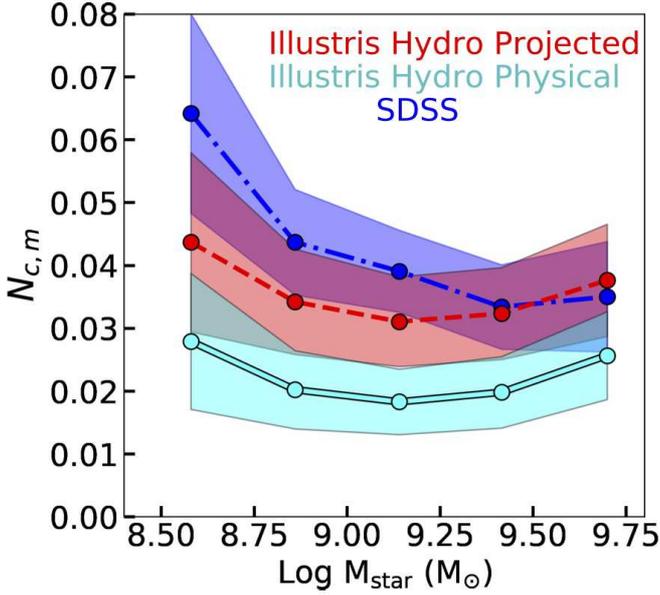}
 \end{center}
 \caption{\label{fig:Group} 
 The mean number of companions is plotted per stellar mass bin, $N_{c,m}$. This represents the number of dwarf companions (of any mass within the defined dwarf mass range) per dwarf in 
 the listed stellar mass bin, normalized by the total
  number of isolated dwarf galaxies in that stellar mass bin. 
  Mass bins are selected as in Fig.~\ref{fig:GalMass}. 
 Mock multiples
  are selected based on either projected separations
 ({\it Projected}; red, dash-dotted) or by first requiring that the dwarfs have 3D separations less than 300 kpc  ({\it Physical}; cyan, solid).
 The observed {\it SDSS} results are denoted in blue (dash-dotted).  
   1$\sigma$ errors are indicated by the shaded regions. 
 At the highest mass bin there is good agreement among all catalogs.  
 At the lower mass bins the {\it Illustris Hydro Physical}  
has a lower rate of multiples than {\it SDSS}, whereas {\it Illustris Hydro Projected Projected}  is a better match.
Note that the higher $N_c$ in the lowest mass bin in the {\it SDSS} catalog may be a result of sensitivity bias towards 
galaxies that are bluer in color, such as interacting dwarf pairs (there are only 130 dwarfs in that lowest mass bin).
  } 
\end{figure}

Fig.~\ref{fig:Group} illustrates the mean number of companions per dwarf per stellar mass bin,  $N_{c,m}$ (i.e. $N_c$ per stellar mass bin). 

$N_{c,m}$ computed for the {\it SDSS} catalog (blue) decreases from the lowest mass bin
towards higher masses.
 The mock catalogs show qualitatively the same behaviour, with the best agreement with {\it SDSS} seen in {\it Illustris Hydro Projected}.

Note that the higher fraction of companions indicated by the {\it SDSS} sample in the lowest mass bin may be a result of sensitivity limits
being biased towards blue colors. Indeed, there are only $\sim$130 dwarfs in the {\it SDSS} sample in the lowest mass bin of Fig.~\ref{fig:Group}. 
\citet{Stierwalt15} showed that dwarfs in pairs have higher SFRs, and 
thus bluer colors on average (see also Fig.~\ref{fig:Mendel}). This suggests that dwarf multiples are easier to identify at the low mass end 
than their non-interacting counterparts. Such observational bias would not be reflected in our mock dwarf selection criterion, where
only limits on stellar mass, not color, were implemented.

\begin{figure}
\begin{center}
\mbox{\includegraphics[width=3.5in, clip=true, trim=0 0 0 0]{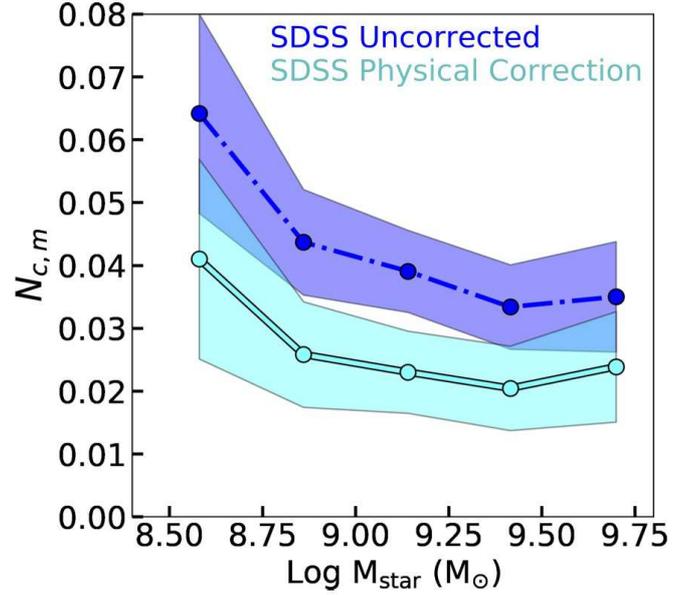}}
 \end{center}
 \caption{\label{fig:GroupCorr} The original {\it SDSS} results for $N_{c,m}$ (blue, dash-dotted) are corrected for projection effects using 
 the difference between $N_{c,m}$ for mock multiples selected using the {\it Illustris Hydro Projected}
 and  {\it Illustris Hydro Physical} criteria from Fig.~\ref{fig:Group}. The correction (cyan, solid line) is on average 30-40\% per mass bin. 
 We caution that the lowest mass bin for the {\it SDSS} results might overestimate the dwarf multiple fraction as {\it SDSS} sensitivity is biased to bluer
 galaxies. Interacting dwarfs are typically blue, making them easier to detect over their quiescent counterparts.
}
\end{figure}

We use the fractional difference 
of $N_{c,m}$ between the {\it Illustris Hydro Physical} and {\it Illustris Hydro Projected} dwarf multiples in each mass bin 
 to correct the {\it SDSS} results of Fig.~\ref{fig:Group} for projection effects.  
 The result, plotted in Fig.~\ref{fig:GroupCorr},
 illustrates the predicted true $N_{c,m}$, if contaminants owing to projection effects were removed.

\subsection{Frequency of Pairs and Groups} 
\label{Freq:Pairs}

We define N$_1$ as the percentage of dwarfs with only one companion (a pair), N$_2$ as having 
only two companions (a triple) and N$_3$ as having 3 companions (a quad).  In practice we do not 
find higher order multiples in the catalogs. Note that with this definition, 
a true triple system, for example, would be counted 3 times, as there are three dwarfs that each have two companions.  
Results for all catalogs are summarized in Table~\ref{table:Np}. 

Roughly 3.2\% of mock dwarfs
in {\it Illustris Hydro Projected} are in a pair, meaning they have one companion within the stellar mass 
range afforded by our adopted sensitivity limits.  This is within the 1$\sigma$ errors of the {\it SDSS} value 
of $3.5 \pm 0.3$\%.  This means that out of a sample of 10,000 isolated dwarfs, 
$\sim$150 unique pairs should exist in the mock sample and 
$\sim$175 in {\it SDSS}.

However, when mock dwarf multiples are required to have 3D separations $< 300$ kpc ({\it Illustris Hydro Physical}) the fraction 
of mock pairs drops by $\sim$40\%.
In other words, out of a sample of 10,000 isolated dwarfs, while there 
should be 175 projected pairs in {\it SDSS}, only $\sim$105 are cosmologically expected to have physical separations less than 300 kpc.

We find that very few dwarfs are found in a triple using any selection criteria in any catalog ($< 0.2$\%). 
Again, taking a sample of 10,000 isolated dwarfs, this fraction implies that at most $\sim$20 dwarfs  have two companions, 
yielding 20/3 $\sim$7 unique projected triples.  

In the {\it Illustris Hydro Physical} catalog of multiples, not all sightlines through the simulation volume yield triples.
Only $\sim$70\% of the 500 realizations of the {\it Illustris Hydro} volume yield any triples. 
Given the rarity of such configurations it is unsurprising
that they are not identified in every realization of the catalogs.  Results quoted in Table~\ref{table:Np} 
are averaged over only sightlines where triples were identified.

\begin{table*}
\centering
\caption{{\bf Percentage of Dwarfs with a Given Number of Companions}
\newline
{\it Columns 3, 4 and 5)} $N_1$ indicates percentage of pairs, $N_2$ triples and $N_3$ quads.  Note that very few sightlines through the {\it Illustris} volume 
host mock quads (only 2-7\% of the 500 realizations of the simulation volume).
  The same is true for mock triples in {\it Illustris Hydro Physical}  (50\% of volume realizations host a triple). But  
 mock triples are identified in 466 of 500 sightlines {\it Illustris Hydro Projected} and 475 of 500 realizations of the SDSS catalog.
 Quoted is the fraction of mock dwarfs in such configurations averaged over all 500 catalog realizations, regardless of whether a quad or triple
 was identified. 
{\it The Last Row)} indicates, for a sample of 10,000 isolated dwarfs, the number of expected {\it unique} {\it Projected}
 pairs, triples and quads, based on the largest percentage listed in the rows above (see text in Section~\ref{sec:FreqPairs}).}
\label{table:Np}
\begin{tabular}{|c|c|c|c|c|}
\hline\hline
Catalog  & Method & $N_1$  & $N_2$ & $N_3$ \\
\hline
SDSS &   Projected   &   3.5 $\pm$ 0.3 &  0.21 $\pm$ 0.08  &  none \\
Illustris Hydro  &  Projected &  3.2 $\pm$ 0.5 &  0.10 $\pm$ 0.08  & 0.004 $\pm$ 0.003    \\
 Illustris Hydro  & Physical &  2.0 $\pm$ 0.4 &  0.08 $\pm$  0.06  & 0.06 $\pm$ 0.03  \\
\hline
10,000 Dwarfs & Projected, Max &  $\sim$160 Pairs & $\sim$3 Triples &  $\lesssim$1 Quad \\ 
\hline
\end{tabular}
\end{table*}

Even fewer realizations of the mock catalogs (7\% in {\it Illustris Hydro Projected} and 2\% in {\it Illustris Hydro Physical}) yield dwarfs with 3 companions (quads). 
It is thus reasonable that none are identified in {\it SDSS}. Of those 2-7\% of realizations, we find only
 0.06\% of mock dwarfs with 3 companions. As such, for a sample of 10,000 isolated dwarfs, it is cosmologically 
 expected that, at 
 best, one may find one quad within our adopted mass constraints. Given that similar statistics are found using the Physical criteria, if a quad is identified under the same criteria as that adopted here, it is likely to be real.

We conclude that groups with more than 3 dwarf galaxy members
($2 \times 10^8$ M$_\odot  < $ M$_{\rm star} < 5 \times 10^9$ M$_\odot$) with angular separations $> 55 \arcsec$, projected separations r$_p  < $ 150 kpc, and relative velocities $<$ 150 km/s are very rare ($< 0.1$\%) both  
cosmologically and observationally at low redshift (z $<$ 0.025).    
 

\subsection {Predictions for Next Generation Surveys: Dwarf Multiples with M$_{\rm star} > 2 \times 10^8$ M$_\odot$}
\label{sec:LSST}

We create predictions for future photometric and spectroscopic surveys that are expected to be complete to stellar 
masses as low as M$_{\rm star} = 2\times 10^8$ M$_\odot$ within a (100 Mpc)$^3$ volume, such as LSST, DESI, etc. We do this by calibrating the 
{\it SDSS} multiples for completeness and projection effects using ``complete'' versions of the  mock catalogs. 

We remove the observational sensitivity limits (Fig.~\ref{fig:Mendel}) in the mock dwarf selection criteria. Instead, we select all dwarfs
with stellar masses M$_{\rm star} > 2 \times 10^8$ M$_\odot$ in {\it Illustris Hydro}  (where
stellar mass is determined as in Section~\ref{sec:MockMass}).   Note that the {\it SDSS} catalog 
is left unchanged. The new mock catalog is called {\it Illustris Hydro Complete}.

\begin{figure}
\begin{center}
\includegraphics[width=3.7in]{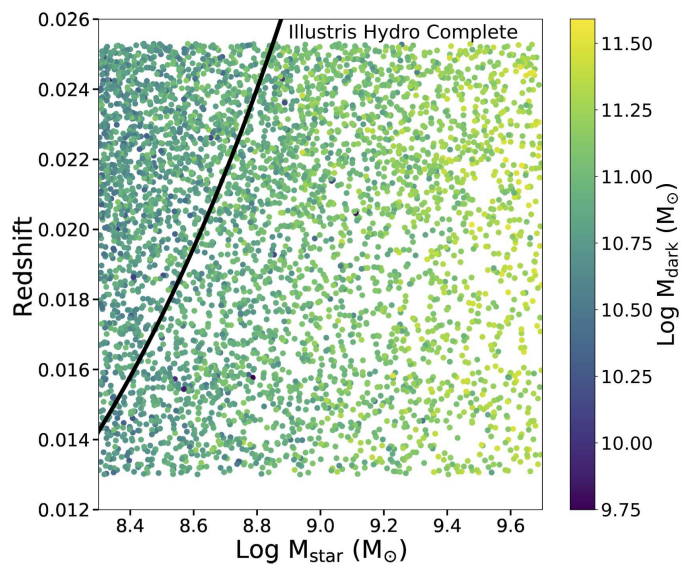}
\end{center}
 \caption{\label{fig:GalMassNoSDSS} 
Distribution of stellar mass for mock isolated dwarfs in the {\it Illustris Hydro Complete} catalog as a function 
 of redshift for one representative realization of the simulation volume. 
 This is the same plot as the right panel of Fig.~\ref{fig:Mendel}, but the observational sensitivity limits
 (black line) are no longer enforced for the mock catalog. Instead,
  mock dwarfs are required to have a hard lower
 stellar mass limit of M$_{\rm star} > 2 \times 10^8$ M$_\odot$.}
 \end{figure}

Fig.~\ref{fig:GalMassNoSDSS} illustrates the distribution of stellar mass for the mock isolated dwarfs in the
 {\it Illustris Hydro Complete} catalog
as a function of redshift. Mock dwarfs are color coded by their halo mass at z=0.  Plotted is one representative realization of the simulation 
volume.  
 
\begin{figure}
\begin{center}
\mbox{\includegraphics[width=3.5in]{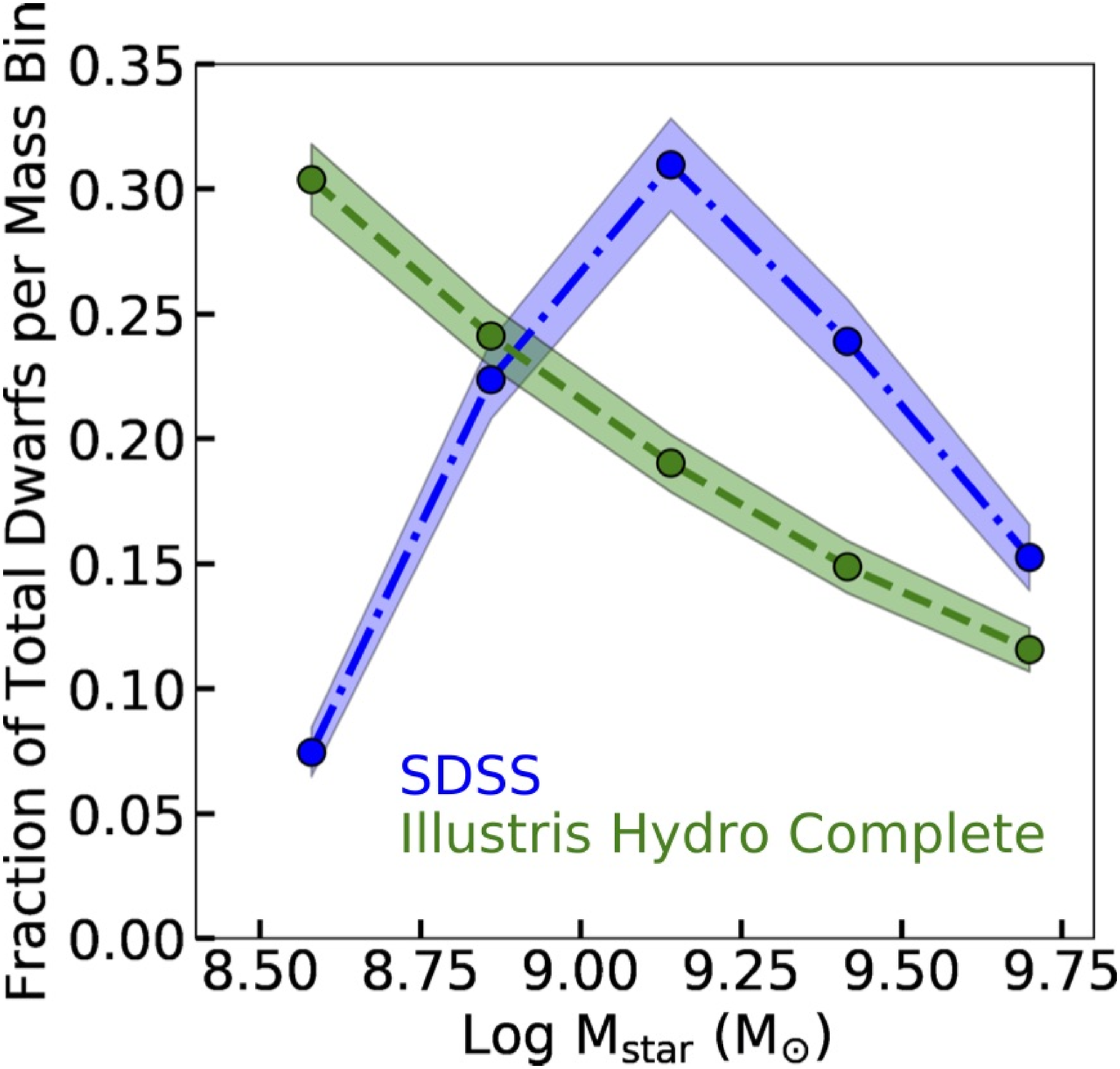}}
 \end{center}
 \caption{\label{fig:GalMassNoSDSS2} 
  The fraction of isolated dwarf galaxies per stellar mass bin, relative to the total number of 
 dwarf galaxies 
 in each respective catalog. This is similar to Fig.~\ref{fig:GalMass}, but with observational sensitivity 
 limits removed for the mock catalogs (corresponding to Fig.~\ref{fig:GalMassNoSDSS}). The {\it SDSS} catalog is unchanged relative to Fig.~\ref{fig:GalMass} (blue; dash-dotted).
    The mock galaxy results for {\it Illustris Hydro Complete} are plotted in green.  2$\sigma$ errors are indicated by the shaded regions.
 Without sensitivity limits, the fraction of mock dwarfs grows sharply towards lower masses. } 
 \end{figure}

Fig.~\ref{fig:GalMassNoSDSS2} illustrates the corresponding fraction of dwarfs per stellar mass bin for 
 {\it Illustris Hydro Complete} and the original {\it SDSS} results (where observational 
sensitivity limits are still enforced). 
The mock dwarf fraction increases towards lower masses, as expected.

We use the {\it Illustris Hydro Complete} catalog to compute the average number of companions per mock dwarf galaxy
($N_c$) using the same {\it Projected} and {\it Physical} criteria listed 
in Section~\ref{sec:SelectionP}. Results are listed in Table~\ref{table:NcNoSDSS}, averaged over all 500 realizations of each catalog. 
The fraction of mock dwarfs with companions 
has predictably increased in each mock catalog. A discrepancy 
remains between the {\it Illustris Hydro Complete Projected} and {\it Illustris Hydro Complete Physical} catalogs, again indicating contamination from projection effects. 

We calibrate the {\it SDSS} dwarf catalog to account for survey incompleteness using the fractional difference between the {\it Illustris Hydro Projected} and {\it Illustris Hydro Complete Projected} results (denoted as $f$).
The difference in $N_c$  
(Table~\ref{table:NcNoSDSS}/Table~\ref{table:Nc}) for the mock {\it Projected} catalog is $f \sim$1.4. This increases predictions for a complete observational survey to $N_c\sim$0.06 (last row of Table~\ref{table:NcNoSDSS}).

If we further include a correction for projection effects, we expect the number of multiples to decrease by $\sim$40\% (e.g. Fig.~\ref{fig:GroupCorr}), yielding $N_c \sim 0.04$.

\begin{table}
\centering
\caption{{\bf Average $N_c$ using Catalogs Complete to M$_{\rm star} > 2 \times 10^8$ M$_\odot$}
\newline 
Same as Table~\ref{table:Nc} except that a stellar mass floor of M$_{\rm star} > 2 \times 10^8$ M$_\odot$ 
has been applied to the mock catalogs. The {\it SDSS} result is unchanged and is listed for reference.
{\it Column 4)}  $f$ indicates the fractional difference in $N_c$  between each complete catalog 
and their incomplete counterpart from Table~\ref{table:Nc}.
{\it The Last Row} is the expected $N_c$ for projected multiples in {\it SDSS}
if it were complete to  
M$_{\rm star} > 2 \times 10^8$ M$_\odot$.
 } 
\label{table:NcNoSDSS}
\begin{tabular}{ |c|c|c|c| }
\hline\hline
Catalog  & Method & $N_c$  &  $f$\\
\hline
SDSS &    Projected  &   0.039 $\pm$ 0.003  & N/A \\
 Illustris Hydro Complete  &  Projected &  0.048 $\pm$ 0.005  & 1.4  \\ 
Illustris Hydro Complete  &  Physical & 0.032 $\pm$  0.004  &  $\sim$1.5  \\  
\hline
SDSS-Complete Correction &   Projected &  {\bf $\sim$0.06}  & 1.4 \\ 
\hline\hline
\end{tabular}
\end{table}

\begin{figure}
\begin{center}
\mbox{\includegraphics[width=3.5in]{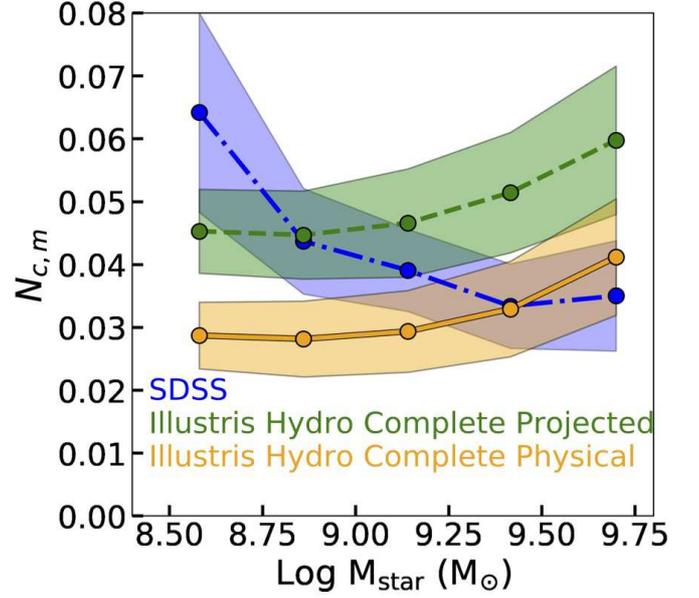}
}
 \end{center}
 \caption{\label{fig:GroupNoSDSS} Same as Fig.~\ref{fig:Group} except that  
 the sensitivity limits have been removed for the mock catalogs.  
 Instead a stellar mass floor of M$_{\rm star} > 2 \times 10^8$ M$_\odot$ is applied to the mock catalogs (orange/solid and green/dashed lines). 
 The {\it SDSS} catalog result (blue; dashed-dotted) is unchanged. 
In both mock-complete catalogs, $N_c$ decreases steadily towards lower mass bins, as expected given the increased sample size
at low masses. Correspondingly, $N_c$ increases at high masses 
relative to the incomplete catalogs in Fig.~\ref{fig:Group}. 
  } 
 \end{figure}

Fig.~\ref{fig:GroupNoSDSS} plots the mean number of companions per mass bin,
 $N_{c,m}$, using both the {\it Illustris Hydro Complete Physical} and {\it Illustris Hydro Complete Projected} catalogs.  This is similar to Fig.~\ref{fig:Group}, but now using the complete mock catalogs.
The {\it SDSS} results (blue) are unchanged, reflecting the original sensitivity limits of the survey.  
In both mock catalogs, 
$N_{c,m}$ increases for the most massive dwarf bins and decreases for low mass bins. This is because the number of mock dwarfs in the lowest mass bins has increased substantially (Fig.~\ref{fig:GalMassNoSDSS2}).
This result supports the hypothesis that the upturn in the {\it SDSS} results at low mass bins is 
likely a result of survey incompleteness.

\begin{figure}
\begin{center}
\mbox{\includegraphics[width=3.5in]{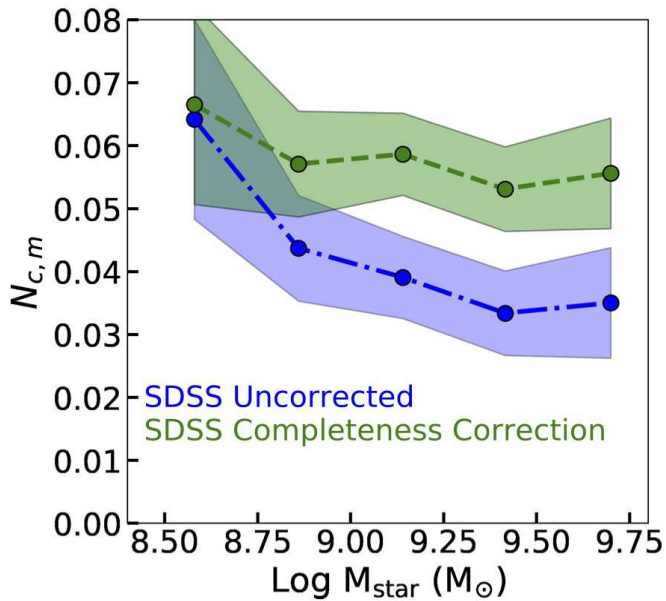}
}
 \end{center}
 \caption{\label{fig:GroupNoSDSSCal}  
 As in Fig.~\ref{fig:Group}, we plot $N_{c,m}$, the average fractional number of dwarf companions per stellar mass bin. 
The original {\it SDSS} catalog results are plotted in blue (dashed-dotted) and corrected for completeness (green; dashed). 
Corrections are made by multiplying the original {\it SDSS} results by
 the fractional change in $N_{c,m}$ computed for the {\it Illustris Hydro Projected} catalogs with or without 
 the observational sensitivity limits (i.e. the ratio of the mock catalogs in Figures~\ref{fig:Group} and \ref{fig:GroupNoSDSS}). 
The resulting {\it SDSS Completeness Correction} catalog is relatively flat across all stellar mass bins at an average value of $N_{c,m} \sim 0.06$, in agreement 
with the average results in Table~\ref{table:NcNoSDSS}. 
Note that results for the lowest mass bin should be viewed cautiously as their may be a bias in the {\it SDSS} catalog towards 
preferentially identifying 
 multiples owing to their bluer color. The cosmological samples suggest instead that the fraction 
 should decrease at lower masses (see Fig.~\ref{fig:GroupNoSDSS}) for the adopted catalog mass range.  
  } 
 \end{figure}

We now calibrate the {\it SDSS} results to predict  
 $N_{c,m}$ for future observational  surveys that will be complete to low masses.  We compute the fractional change in 
 $N_{c,m}$ between the {\it Illustris Hydro Projected} catalogs {\it without} the sensitivity limits (Fig.~\ref{fig:GroupNoSDSS}) and
  {\it with} them (Fig.~\ref{fig:Group}) in each mass bin. 
 In Fig.~\ref{fig:GroupNoSDSSCal}
 we utilize this fractional change to correct the {\it SDSS} $N_{c,m}$ for completeness (green; dashed).   
 
The {\it SDSS Completeness Correction} (green; dashed) results for $N_{c,m}$ are relatively flat across all mass bins at an average value of $\sim$0.06. This is the prediction for the mean number of companions per dwarf galaxy in future surveys that are complete to stellar masses of $2 \times 10^8$ M$_\odot$ over a $\sim$(100 Mpc)$^3$ volume.

\section{Discussion}
\label{sec:Discussion}

In the following, we use the {\it SDSS} and mock dwarf-multiple catalogs to connect with existing 
studies of dwarf-dwarf interactions in the literature. We first
quantify the frequency of ``Major Pairs''  compared to studies 
of the ``Major Merger'' rate of dwarf galaxies (Section~\ref{sec:MajMerger}). We next quantify the observed and cosmologically expected
frequency of analogues of the Magellanic Clouds 
in the field,
comparing our
results to the known frequency of such analogues near Milky Way-type hosts (Section~\ref{sec:LMC}). 
Finally we place the recently discovered set of 7 projected dwarf groups from the TiNy Titans ({\it TNT}) survey
\citep{Stierwalt17} within the context of our study (Section~\ref{sec:TNT}).

\subsection{``Major Pair'' Fraction, M$_{\rm S,star}/$M$_{\rm P,star} > 1/4$ } 
\label{sec:MajMerger}

Galaxy interactions are more destructive as the mass ratio of the interacting systems increases.
We define destructive ``Major Pairs'' as dwarf galaxy 
pairs with stellar mass ratios of M$_{\rm S,star}/$M$_{\rm P,star} > 1/4$. S,star refers to stellar mass of the 2nd most massive member (Secondary) and P,star to that of the most massive member (Primary).

\begin{figure}
\begin{center}
\includegraphics[width=3.5in]{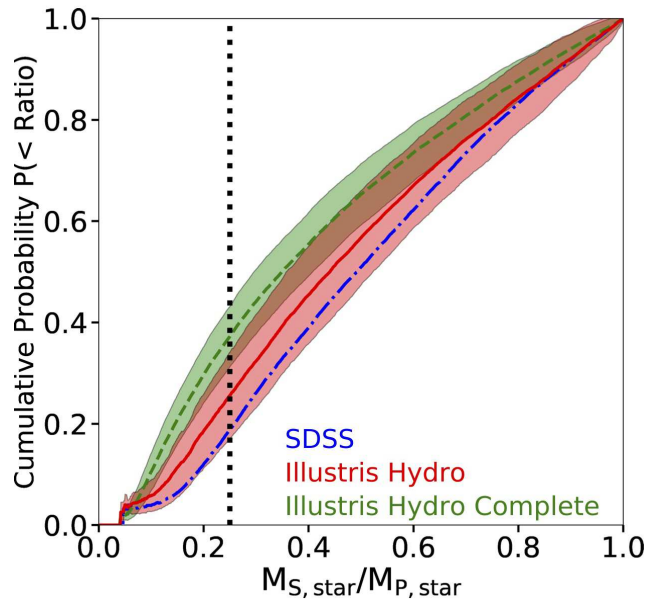}
 \end{center}
 \caption{\label{fig:Ratio} 
The cumulative probability distribution of the stellar mass ratio of Secondaries/Primaries (M$_{\rm S,star}/$M$_{\rm P,star}$) is plotted for multiples in the {\it SDSS} catalog (blue; dash-dotted) and {\it Illustris Hydro Physical} catalog (red; solid line). Results for {\it Illustris Hydro Projected} are similar. 
The shaded regions indicate 1$\sigma$ deviation of the mean, computed over 500 realizations of each catalog (SDSS errors are encompassed within the red shaded region).  70-85\% of all pairs in the sensitivity limited catalogs are ``Major Pairs'', defined as having M$_{\rm S,star}/$M$_{\rm P,star} > 1/4$ (to the right of the limit denoted by the black dotted line). 
Results for {\it Illustris Hydro Physical Complete} (green; dashed line), indicate that for catalogs that are complete to M$_{\rm star}= 2\times10^8$ M$_\odot$, ``Major Pairs'' are cosmologically expected to comprise 60-70\% of the catalog. } 
 \end{figure}

Fig.~\ref{fig:Ratio} illustrates the cumulative probability distribution of stellar mass ratios of dwarf pairs identified in the {\it SDSS} 
and {\it Illustris Hydro Physical} catalogs. Results for the {\it Illustris Hydro Projected} catalog are similar; the stellar mass ratio distribution is unaffected by projection effects. 
We find that, within our adopted stellar mass and sensitivity limits, 70-85\% of dwarf pairs identified in all catalogs are ``Major Pairs''.  

The large fraction of high mass ratio encounters is partly a result of our sensitivity and mass limits. Results for the {\it Illustris Hydro Physical Complete} catalog is plotted in Fig.~\ref{fig:Ratio}, where the sensitivity limits are dropped (as in Section~\ref{sec:LSST}). 
Now 60-70\% of dwarf pairs are ``Major Pairs''. This fraction is certain to decrease as the mass limits of our surveys are lowered.
As such, the stellar mass ratio distribution should not be confused/compared with a ``complete'' mass ratio distribution (e.g. Figs. 6 and 7 from \citet{Rodriguez15}, bottom panels).

\begin{figure}
\begin{center}
\mbox{\includegraphics[width=3.5in]{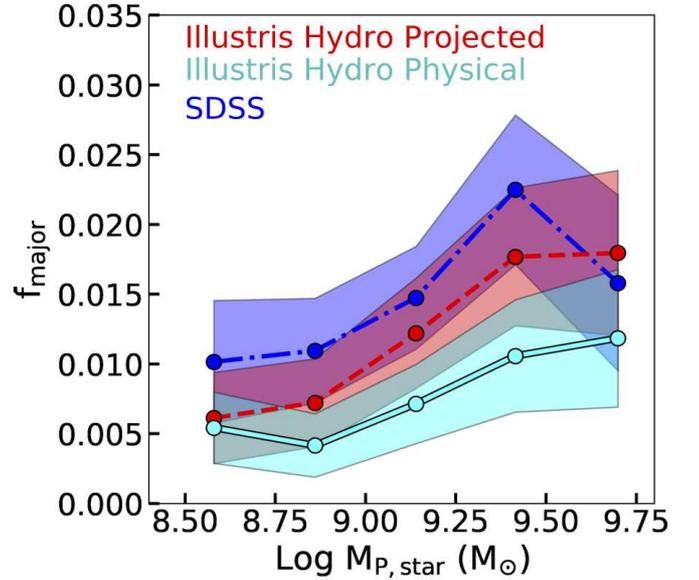}
}
 \end{center}
 \caption{\label{fig:MajMerger} 
  The Number of Primary dwarf galaxies that have a Secondary with a stellar mass ratio of M$_{\rm S,star}/$M$_{\rm P,star} > 1/4$ (``Major Pairs'')
 per Primary stellar mass bin, normalized by the total
  number of isolated dwarf galaxies in that stellar mass bin (f$_{\rm major}$). Mass bins are defined as in Fig.~\ref{fig:Group}.
  Results for the {\it Illustris Hydro Projected} catalog are in red (dashed line) and {\it Illustris Hydro Physical}  in  cyan (solid line).
   {\it SDSS} results are shown in blue (dash-dotted). Shaded regions
   indicate 1$\sigma$ errors and  lines the mean values. 
 {\it Illustris Hydro Projected} results agree best with the {\it SDSS} catalog.
 The ``Major Pair'' fraction at high mass (Log M$_{\rm star} > 9.2$) is 1.2-2.5\%; this is a robust result, 
  as all catalogs are roughly complete at these masses. 
  } 
 \end{figure}

Fig.~\ref{fig:MajMerger} plots f$_{\rm major}$, the number of all dwarf Primaries
 in a given stellar mass bin that have a Secondary dwarf galaxy companion with a stellar mass ratio of M$_{\rm S,star}/$M$_{\rm P,star} > 1/4$, normalized by the total number of dwarfs in the bin.
Good agreement is found between the {\it SDSS} and {\it Illustris Hydro Projected} catalogs: the
``Major Pair'' fraction  
peaks at high Primary masses at a value of $f_{\rm major} \sim 0.02$ and steadily declines towards lower Primary masses, reaching a value of $f_{\rm major} \sim 0.01$.
The {\it Illustris Hydro Physical} catalog shows the same qualitative behaviour, but with lower values.

\begin{figure*}
\begin{center}
\mbox{\includegraphics[width=3.55in]{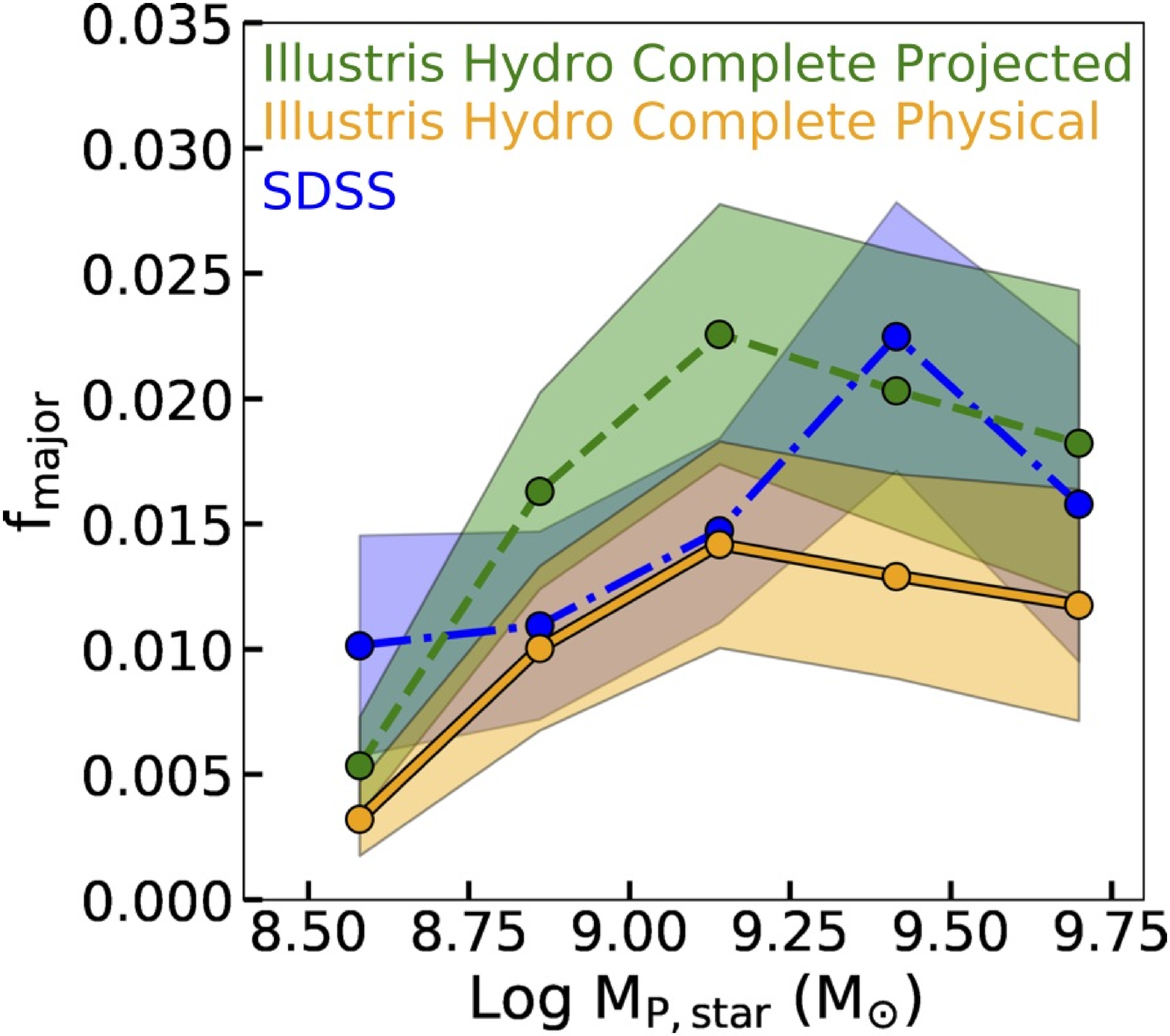}
\includegraphics[width=3.3in, clip=true, trim=0.89in 0in 0in 0in]{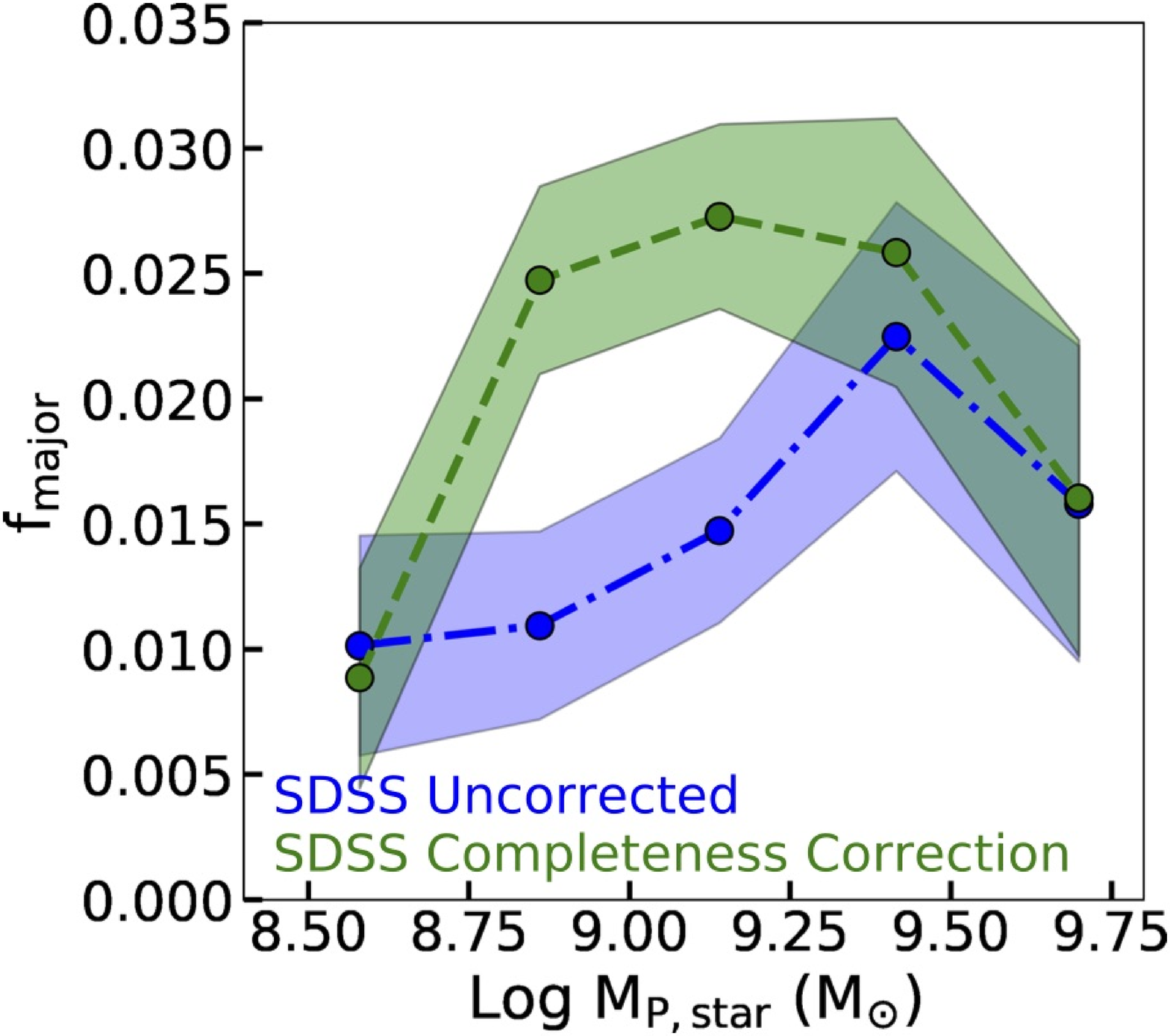}
}
 \end{center}
 \caption{\label{fig:MajMergerNoSDSS}  {\it Left:} Same as Fig.~\ref{fig:MajMerger} except that  
 the sensitivity limits have been removed for the mock catalogs.  Instead we require only a stellar mass floor of $2 \times 10^8$ M$_\odot$. {\it SDSS} results are unchanged (blue, dash-dotted).  Note that the rapid drop in the lowest Primary mass bins results from a deficit of lower mass dwarfs in the catalogs.
 As expected, $f_{\rm major}$ does not change significantly for the most massive bin, which is roughly complete for all catalogs. 
In a complete catalog, 1.8-2.5\% of dwarfs with stellar masses in excess of $10^9$ M$_\odot$ are cosmologically expected to be found in a projected ``Major Pair''. 
Accounting for pairs within a 300 kpc separation ({\it Illustris Hydro Complete Physical}) reduces these values  by a factor of $\sim$1.7. 
{\it Right:}  We correct the {\it SDSS} results for completeness (green, dashed line) using the fractional difference between the {\it Illustris Hydro Complete Projected} (left panel) and {\it Illustris Hydro Projected} (Fig.~\ref{fig:MajMerger}) results per mass bin.  The blue, dash dotted line is the original {\it SDSS} results. 
Surveys like 
LSST are predicted to find a dwarf ``Major Pair'' fraction that increases with decreasing Primary stellar mass.  The
leveling off at $\sim10^9$ M$_\odot$ likely owes to the lower stellar mass limit of the catalogs.
  } 
 \end{figure*}

In Fig.~\ref{fig:MajMergerNoSDSS}
we assess how much incompleteness has affected our results using the {\it Illustris Hydro Complete Projected} and {\it Physical} catalogs, where
sensitivity limits are removed and only a stellar mass floor of $2\times 10^8$ M$_\odot$ is enforced (see Section~\ref{sec:LSST}). Both  complete mock catalogs show the same qualitative behaviour. 
The complete mock ``Major Pair'' fraction 
stays roughly flat between Primary stellar masses $\sim \log$(M$_{\rm star}$) = 9.1-9.4 M$_\odot$ at an average value of $f_{\rm major} \sim 0.02$ for {\it Illustris Hydro Complete Projected} and  $f_{\rm major}\sim 0.012$ for {\it Illustris Hydro Complete Physical}.
 This result is robust, 
  as all catalogs are roughly complete at those masses.

Note that the sharp drop off at lower stellar
mass bins is a direct result of the lower stellar mass limit, and is not physical.  

We utilize these results to calibrate the {\it SDSS} ``Major Pair'' fraction for completeness using the fractional change in the {\it Illustris Hydro Projected}
``Major Pair'' fraction with (Fig.~\ref{fig:MajMerger}) and without (Fig.~\ref{fig:MajMergerNoSDSS}) the {\it SDSS} sensitivity limits. 
The result is plotted in the right panel of Fig.~\ref{fig:MajMergerNoSDSS}: the ``Major Pair'' fraction is cosmologically expected 
to continuously increase towards lower Primary stellar masses (within Log M$_{\rm P,star} = 9.0-9.75$).  
At lower Primary masses, the fraction levels off at $f_{\rm major} \sim$0.027, but this leveling off likely due to the lower stellar mass limits. This behaviour is qualitatively similar to that found by  \citet{Casteels14}, who defined ``Major Mergers'' based on {\it galaxy pair} separation and asymmetry 
 in the GAMA survey \citep{Driver09,Driver11}. Note, however that their study accounts for dwarfs in all environments, whereas we focus only on isolated systems. 
 
 \citet{Sales13} find that the {\it average} satellite abundance is largely independent of Primary mass for galaxies with M$_{\rm star} < 10^{10}$ M$_\odot$). Here, we find that that the frequency of {\it isolated} ``Major Pairs'' appears to be a strong function of Primary stellar mass. But this apparent conflict is not surprising, given the rarity of ``Major Pair'' configurations.

 \citet{Rodriguez15} derived the galaxy-galaxy ``Major Merger'' rate using merger histories computed with the 
SUBLINK code and the {\it Illustris Hydro} simulation. This is distinct from our method, as they count coalesced systems, whereas we 
identify dwarf pairs as distinct subhaloes\footnote{Note that deriving a ``Major Merger'' rate using our pair fractions is difficult as the orbital timescales 
of interacting dwarfs are unknown. From studies of the Magellanic Clouds (stellar mass ratio M$_{\rm S,star}$/M$_{\rm P,star}$ = 0.1) the time of coalescence can be 
very long \citep[$>$ 6 Gyr,][ see also,]{Pearson18, Besla16}. We defer calculations of merger rates 
to a future study where we will study the kinematics of our dwarf multiples and orbital timescales in detail.}. 
\citet{Rodriguez15} find that the ``Major Merger'' rate steadily declines with decreasing Primary stellar mass, whereas we find the opposite behaviour for ``Major Pairs'' at z$\sim$0. 
This suggests that the galaxy merger timescale is both a function of redshift \citep{Snyder17} and galaxy mass, even at dwarf mass scales \citep[see also][]{Kitzbichler08,Conselice06}. Note that we consider only isolated ``Major Pairs'', whereas \citet{Rodriguez15} measure the ``global Major Merger'' rate, regardless of environment. It is plausible that environmental effects (e.g. the tidal field of the host) could impact the merger rate of dwarf pairs after capture by a more massive host.

Regardless, our result indicates that long-term, repetitive dwarf-dwarf interactions may play an important role in the SFHs, the origin of starbursts and gas removal processes in isolated dwarf galaxies. 
  Because dwarf galaxies in the field have significant gas fractions \citep[$f_{\rm gas} > 50\%$,][]{Bradford15,Geha12}, tidal interactions between dwarfs are more likely to remove gas than stars. The creation of extended, long-lived gas tidal structures around dwarf pairs will affect the nature of their circumgalactic medium and impact their baryon cycles, both in isolation and after accretion into a more massive environment \citep{Pearson18}.

\subsection{The Frequency of Isolated LMC \& SMC Analogs}
\label{sec:LMC}

The closest example of dwarf-dwarf galactic pre-processing
at work
is the Magellanic System. The interacting dwarf pair, 
the Large Magellanic Cloud (LMC) and 
Small Magellanic Cloud (SMC), 
are enveloped by a massive \citep[$2\times 10^9$  M$_\odot$;][]{Fox14} complex of gas called the Magellanic Stream, Bridge and Leading Arm
\citep{Mathewson74,Putman03,Nidever10}. This extended gaseous complex is primarily created by the tidal interaction between the two dwarfs \citep{Besla10,Diaz11,Besla12,Besla13,Guglielmo14,Pardy18}.
Furthermore, orbital solutions indicate that the Clouds are recent interlopers to our Galaxy \citep{Kalli13, Besla07}. This implies that the Magellanic Clouds must have been a relatively isolated, interacting galaxy
pair prior to their capture by the Milky Way. 
This study is designed to determine whether such isolated dwarf pair configurations are expected cosmologically. 

The frequency of analogs of the Magellanic Clouds about massive hosts has been quantified in both cosmological 
simulations and {\it SDSS}, finding good agreement.  
Roughly, 40\% of Milky Way analogs in both cosmological simulations and {\it SDSS} 
host an LMC-type galaxy \citep{Boylan-Kolchin11, Busha11, Tollerud11, Patel17, Robotham12}.
On the other hand, it is rare ($\sim$2-5\%)
to find both an LMC and SMC analog in proximity to a massive host \citep{Liu11,Boylan-Kolchin11, Busha11, Robotham12, Gonzalez13}. The observational statistics become minuscule if one further requires a Milky Way-LMC-SMC analog with clear signs 
of interaction between the two satellites or that the LMC/SMC analogs also be in close proximity to each other \citep{James11,Paudel17}.
However, the cosmologically expected frequency of LMC-SMC analogs (mass, separation, relative velocity) in the {\it field} is unknown. 

From Fig.~\ref{fig:Ratio}, we find that Primaries with companions of stellar mass ratios of 1:5 represent
$\sim 20\%$ of each catalog of dwarf multiples. 
We refine this analysis to identify LMC-SMC analogs with similar separations, 
stellar masses, and mass ratio as the real system.
LMC analogs are defined as dwarfs with stellar masses of $10^9$ M$_\odot  < $ M$_{\rm star} < 5 \times 10^9$ M$_\odot$. 
On average we find $\sim 1700 \pm 200$ dwarfs satisfying these criteria in {\it Illustris Hydro}. The standard deviation results from a combination of scatter in the abundance matching relations and variations in the number of dwarfs in different realizations of the simulation volume.

We define SMC analogs as those dwarfs with stellar mass ratios of 1:15 $<$ LMC:SMC $<$ 1:5 and projected separations of r$_p < 100$ kpc to an LMC analog.  Separation limits are motivated by \citet{Besla12}, who find a best fitting orbit for the 
LMC/SMC with apocenters reaching such distances. 
Examples of the properties for LMC/SMC analogs in one realization of the {\it SDSS} and mock catalogs are plotted in Figures~\ref{fig:DMStars} and ~\ref{fig:RpVSZ} 
where LMC/SMC analogs are marked by symbols with orange outlines. 

For all samples, LMC analogs that host SMC analogs have, on average, present day dark matter subhalo masses a factor of $1.2-1.3$ times more massive than those that do not, in agreement with the recent study by \citet{Shao18}. This result supports high mass models for the total dark matter mass of the LMC when it first entered the Milky Way \citep[][Garavito-Camargo, et al. in prep]{Besla12,Penarrubia16}. 

We further find that $\sim 0.2-0.3\%$ of LMC mass galaxies in the field host an SMC-like companion in both {\it SDSS} and {\it Illustris Hydro Projected} (averaged over 500 realizations of the catalogs). This result is confirmed in {\it Illustris Hydro Physical}, indicating that pairs selected with r$_p < 100$ kpc are robust to projection effects. 
Removing the observational sensitivity limits doubles the frequency of LMC/SMC analogs, but overall we find them to be very rare in the field (cosmologically and observationally).  Of order 1\% of LMC analogs are expected to be found with an SMC-mass companion in a catalog that is complete to SMC
 masses of $2\times 10^8$M$_\odot$.

 \citet{Robotham12} utilized the GAMA Survey to explore the frequency of LMC/SMC analogs as a function of environment.
 They found that $\sim$4.8\% of galaxies with $-19 < M_r < -17$ have a companion in this magnitude range within a projected separation
 of $<$100 kpc and velocity separation of $<$100 km/s (46 out of 1929 galaxies). This is higher than what we find here, but their results consider a wider range of environments.  These authors find a higher probability of finding LMC/SMC analogs in proximity 
of a Local Group analog (within 1 Mpc). Indeed, half (27) of such pairs are within 1 Mpc of an L* galaxy and would fail our isolation criteria, where 
all dwarfs are at least 1.5 Mpc away from an L* type system. This
reconciles our statistics with the GAMA study ($<1$\% of GAMA LMC/SMC analogs would pass our isolation criteria). 

While we find that LMC/SMC configurations are rare today, it is possible that they have been more frequent at earlier times. The good agreement between the mock catalogs and {\it SDSS} indicate that cosmological simulations are reasonable tools to explore the kinematics and frequency of such configurations across cosmic time, an analysis that we defer to a later study.


\subsection { The TNT Dwarf Groups in a Cosmological Context }
\label{sec:TNT}

The TiNy Titans ({\it TNT}) survey of dwarf galaxy pairs identified in {\it SDSS} at low redshift ($0.005 < z < 0.07$)
showed that the SFRs of dwarf pairs increase with decreasing pair separation. 
This suggests that coalescence is not required to induce a burst of star formation. Moreover, the secondary galaxy 
was typically the pair member undergoing the starburst. Together, these results strongly suggest that the average
 dwarf galaxy multiple fractions ($N_c$) that we have determined in 
previous sections, are important to understanding the drivers of starbursts in dwarf galaxies. 
 
\citet{Lee09} find that 6\% of dwarfs within an 11 Mpc volume of the Milky Way are currently star bursting. 
In Section~\ref{sec:LSST}, we found that a comparable fraction ($\sim$4\%) of isolated dwarfs in {\it SDSS} should have a dwarf galaxy companion with angular separations $> 55\arcsec$, projected separations r$_p < 150$ kpc, 
 physical separations $< 300$ kpc, and a relative velocity difference of $<$150 km/s. 

Recently, \citet{Stierwalt17} (hereafter S17) identified 7 isolated groups of dwarfs with 3 members or more at low redshift, including a group 
with 5 members. All dwarf groups are identified to have projected separations less than 80 kpc and most have separations 
greater than 15 kpc. Furthermore, most dwarf members have velocities relative to the Primary that are 
less 200 km/s.  These properties are comparable to our selection criteria for the {\it SDSS} and {\it Illustris Hydro Projected} catalogs of dwarf multiples. 

In Section~\ref{Freq:Pairs} we found it highly improbable to find groups of 4 members or more at low redshift. 
At first glance these results seem at odds with S17. 
However, the dwarf stellar mass range explored in S17 extends to lower masses than we consider in this study. This coupled 
with our restrictive velocity cut of $<$150 km/s reduces 4 of the dwarf groups in S17 to dwarf pairs and makes 3 companionless.  As such, the findings from S17 are consistent with our study, where we find that dwarfs within our mass range 
are more likely to be in pairs, rather than larger multiples.

Unfortunately, because of the mass limits adopted in our study (motivated by the observational sensitivity limits of the {\it SDSS} survey and mass resolution limits of the simulations), we cannot 
state with certainty that the S17 groups containing dwarfs with masses below $\log$(M$_{\rm star}$) =8.3 are consistent with cosmological 
expectations.  Our results do suggest that high speed members ($\Delta$V $>$ 150 km/s) of quads or quints in the dwarf mass range explored in this study are likely projected contaminants. However, if quads are identified to satisfy the position and velocity ($\Delta$V $<$ 150 km/s) constraints of this study, the agreement between the frequency of quad identification in {\it Illustris Hydro Projected} and {\it Illustris Hydro Physical} (see Section~\ref{sec:FreqPairs}) do suggest that they are not chance alignments.


\section{Conclusions}
\label{sec:Conclusions}

The frequency of companions to isolated dwarf galaxies (M$_{\rm star}$ = $2 \times 10^8$ - $5 \times 10^9$ M$_\odot$) 
at low redshift ($0.013 < z < 0.0252$)
is quantified in the {\it SDSS} Legacy spectroscopic catalog (M14) and compared against 
cosmological expectations using
mock catalogs constructed using the {\it Illustris-1} hydrodynamic  cosmological simulation ({\it Illustris Hydro}). We have chosen stellar mass and redshift ranges where the cosmological and observed catalogs can be reasonably compared, accounting for both the resolution of the simulation and the sensitivity limits of {\it SDSS}.

We define dwarf galaxies to be isolated if no Massive Galaxy (M$_{\rm star} > 5 \times 10^9$M$_\odot$) can be identified with both a relative velocity of $\Delta V_{\rm LOS} < 1000$ km/s and a Tidal Index $\Theta > -14.9$ \citep{Karachentsev13}.

Overall we find good agreement between {\it SDSS} and cosmological expectations for the fraction of dwarf multiples (pairs/groups) in isolated environments. Our main results are summarized in the following. 

\begin{itemize} 

\item {\bf There are more dwarfs in the field in the cosmological simulations relative to \emph{SDSS}.}   We confirm results from \citet{Klypin15} that the abundance of mock dwarf galaxies in the field is higher than observed. Using {\it Illustris Hydro}, we find densities $\sim$2.0 times higher than in {\it SDSS} (Table~\ref{table:IsolCounts}). This overabundance of field dwarfs (where environmental effects are negligible) may indicate that many low-surface brightness dwarfs are currently missing from the {\it SDSS} spectroscopic catalog \citep{Danieli18,Greco17}, or may be reconciled with improved subgrid physics models \citep{Wetzel16} and the inclusion of reionization.  Regardless, the solution must be consistent across the dwarf stellar mass range, as a similar fraction of dwarfs are missing in each mass bin, which may be challenging.

\item{\bf The fraction of isolated dwarfs as a function of stellar mass in {\it SDSS} agrees with cosmological expectations,} (see Fig.~\ref{fig:GalMass}). We thus focus our study on the fraction of dwarfs found in a multiple (pairs, triples, quads, etc.). 

\item {\bf The mean number of companions per isolated dwarf is \emph{$N_c \sim 0.04$}.}  We identify dwarf companions based on the following {\it Projected} criteria: 1) an angular separations of $>55\arcsec$ and a projected separation of r$_p < 150$ kpc; 
and 2) a relative line of sight velocity of $\Delta V_{\rm LOS} < 150$ km/s.  Following the methodology of \citet{Patton00} and applying these {\it Projected} criteria, we find the mean number of low mass companions per dwarf to be 
$N_c = 0.039 \pm 0.003$ in {\it SDSS}, which agrees within 1$\sigma$ of the cosmological catalog ($N_c =$ 0.034$\pm$0.005 for {\it Illustris Hydro Projected}). These fractions are certain to be higher if lower stellar mass companions were considered in this study.

\item {\bf $\sim$40\% of isolated dwarf multiples are false associations owing to projection effects.} To assess the degree of contamination from projected pairs with large 3D separations, we add a requirement that all companions have a 3D separation of $r_{3D} <$300 kpc ({\it Illustris Hydro Physical}). 
This reduces $N_c$ by 40\%, indicating a significant contamination fraction exists when selecting dwarf pairs based on projected properties.  We calibrate the {\it SDSS} findings for  contamination from such projection effects, finding a true average fractional number of physical companions per dwarf to
be $N_c = 0.027$.

\item{\bf The majority of isolated dwarf multiples in our mass range are Pairs. Triples are rare and higher order multiples are cosmologically improbable within our adopted redshift, mass, separation and relative velocity limits.}  
Less than $0.2$\% of mock or observed isolated dwarfs are found in a triple.
Most mock triples in our mass range contain projected contaminants;  most high order multiples are no longer identified when dwarf companions are required to have 3D separations $<$ 300 kpc {\it Illustris Hydro Physical}.

\item {\bf The recently discovered \emph{TNT} groups (S17) are reconcilable with cosmological expectations.} Our results do not conflict with the recent discovery of 7 high order dwarf multiples by the {\it TNT} group (S17), as most members of those groups would not satisfy our selection criteria, either owing to their low stellar mass or high relative speed, which reduce the groups to pairs or single dwarfs according to our selection criteria.

\item {\bf We predict the average number of companions per isolated dwarf to be \emph{$N_c \sim 0.06$} in future surveys that are complete to M$_{\rm star} = 2\times 10^8$ M$_\odot$.} 
We remove sensitivity limits to construct mock galaxy catalogs, applying instead a stellar mass floor of M$_{\rm star} = 2 \times 10^8$ M$_\odot$.  This increases the fraction of multiples by a factor of $f \sim 1.4$ in {\it Illustris Projected Hydro} ($N_c = 0.048 \pm 0.005$). 
We utilize this result to calibrate the observed fraction of dwarf multiples in {\it SDSS} for future surveys, finding an expected average fraction of
{\it Projected} companions per dwarf of $N_c =0.06$. 
This value is expected to be roughly constant across the entire dwarf mass range explored in this study.
 Testing these predictions will require a spectroscopic complement to deep photometric imaging surveys (e.g the combination of both LSST and DESI). 

\item {\bf $< 1$\% of isolated LMC analogs have an SMC-like companion.}   For surveys complete to M$_{\rm star} = 2$ $\times$ $10^8$ M$_\odot$ (i.e., the stellar mass of the SMC) out to 100 Mpc, the cosmological catalogs indicate that 1\% of LMC mass analogs are expected to have an SMC-like companion (stellar mass ratio of 1:5-1:15 and r$_p < 100$ kpc). This is in agreement with the {\it SDSS} catalogs and results from the GAMA survey \citep{Robotham12}. We conclude that analogs of the LMC and SMC pair are rare in the field at low z.

\item {\bf The fraction of isolated ``Major Pairs'' increases with decreasing stellar mass in the dwarf regime.} We find good agreement between the fraction of ``Major Pairs'', dwarf pairs with stellar mass ratios $> 1:4$, in {\it SDSS} and {\it Illustris Hydro Projected}  (Fig.~\ref{fig:MajMerger}).  
Correcting the {\it SDSS} catalog for completeness, the ``Major Pair'' 
fraction increases with decreasing stellar mass, from $f_{\rm major} \sim$ 0.015 at $\sim$LMC masses to $f_{\rm major}\sim$0.027 at lower masses. This is the opposite behaviour of the cosmological ``Major Merger'' rate, defined by the coalescence of two galaxies identified by the extracted merger trees for each decendent halo \citep{Rodriguez15}. 
This suggests that the merger timescales for dwarfs are not only redshift-dependent \citep{Snyder17} but also mass-dependent.
\end{itemize}

The good agreement between observations and cosmological expectations for the fraction of dwarf multiples at z=0 indicates that we can reasonably utilize the kinematic properties of cosmological dwarfs multiples to understand their dynamical state and merger timescales across cosmic time.

\section*{Acknowledgements}

We thank Marla Geha, Greg Snyder and Nicolas Garavito-Camargo for useful conversations that have facilitated this work. We also thank the {\it Illustris} collaboration for making their subhalo catalogs and merger trees publicly available.

G.B. acknowledges that this material is based on work supported by the National Science Foundation under grant AST 1714979.
N.K. is supported by NSF CAREER award 1455260.
D.R.P.  acknowledges the support of the Natural Sciences and Engineering Research Council of Canada (NSERC).

The analysis in this study was carried out using the 
the El Gato cluster at the University of Arizona,
which is funded by the National Science Foundation through Grant No. 1228509.

This work has also used catalogs from the
NASA Sloan Atlas and the SDSS. Funding for the NASA Sloan
Atlas has been provided by the NASA Astrophysics Data
Analysis Program (08-ADP08-0072) and the NSF (AST-
1211644).

Funding for SDSS-III has been provided by the Alfred P. Sloan Foundation, the Participating Institutions, the National Science Foundation, and the U.S. Department of Energy Office of Science. The SDSS-III web site is http://www.sdss3.org/.

SDSS-III is managed by the Astrophysical Research Consortium for the Participating Institutions of the SDSS-III Collaboration including the University of Arizona, the Brazilian Participation Group, Brookhaven National Laboratory, Carnegie Mellon University, University of Florida, the French Participation Group, the German Participation Group, Harvard University, the Instituto de Astrofisica de Canarias, the Michigan State/Notre Dame/JINA Participation Group, Johns Hopkins University, Lawrence Berkeley National Laboratory, Max Planck Institute for Astrophysics, Max Planck Institute for Extraterrestrial Physics, New Mexico State University, New York University, Ohio State University, Pennsylvania State University, University of Portsmouth, Princeton University, the Spanish Participation Group, University of Tokyo, University of Utah, Vanderbilt University, University of Virginia, University of Washington, and Yale University.

This research also utilized: IPython \citep{Perez07}, numpy \citep{vanderWalt11} and matplotlib \citep{Hunter07}.



\appendix
\section{Comparison with Illustris-Dark-1}
\label{sec:DarkCompare}

Throughout this study we have utilized the {\it Illustris Hydro} simulation. Here we show that our results are consistent using the dark matter only version of the simulation, {\it Illustris-Dark-1}.

We follow the same methodology as outlined in Section~\ref{sec:MockMass} to assign a  stellar mass to each subhalo in the simulation. The number density of isolated mock dwarfs in {\it Illustris-Dark-1} is $n_{\rm dwarfiso} = 0.0056 \pm 0.0008$, in good agreement with {\it Illustris Hydro} (see Table~\ref{table:IsolCounts}). 

\begin{figure*}
\begin{center}
\mbox{\includegraphics[width=3.0in]{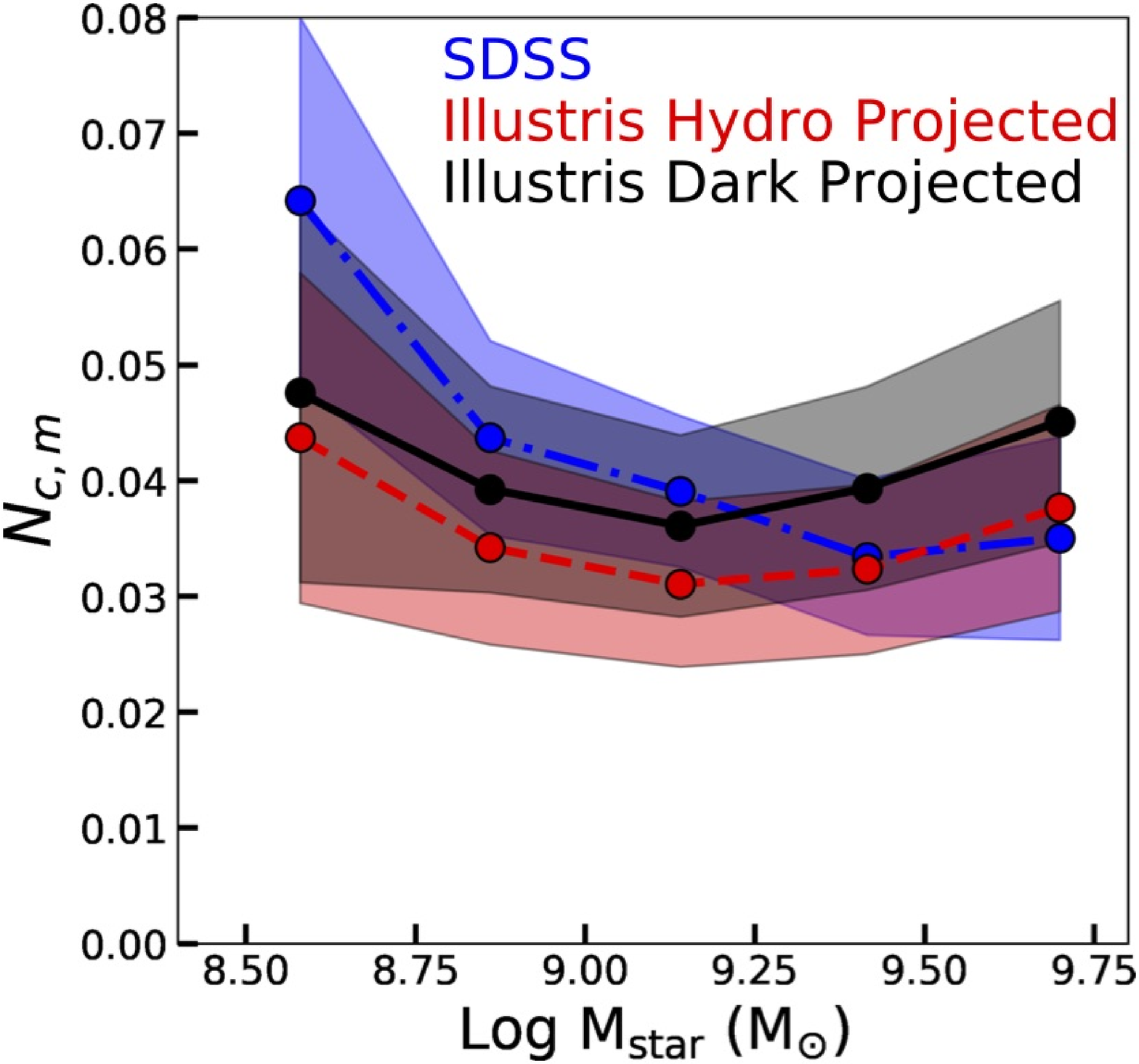}
\includegraphics[width=3.1in]{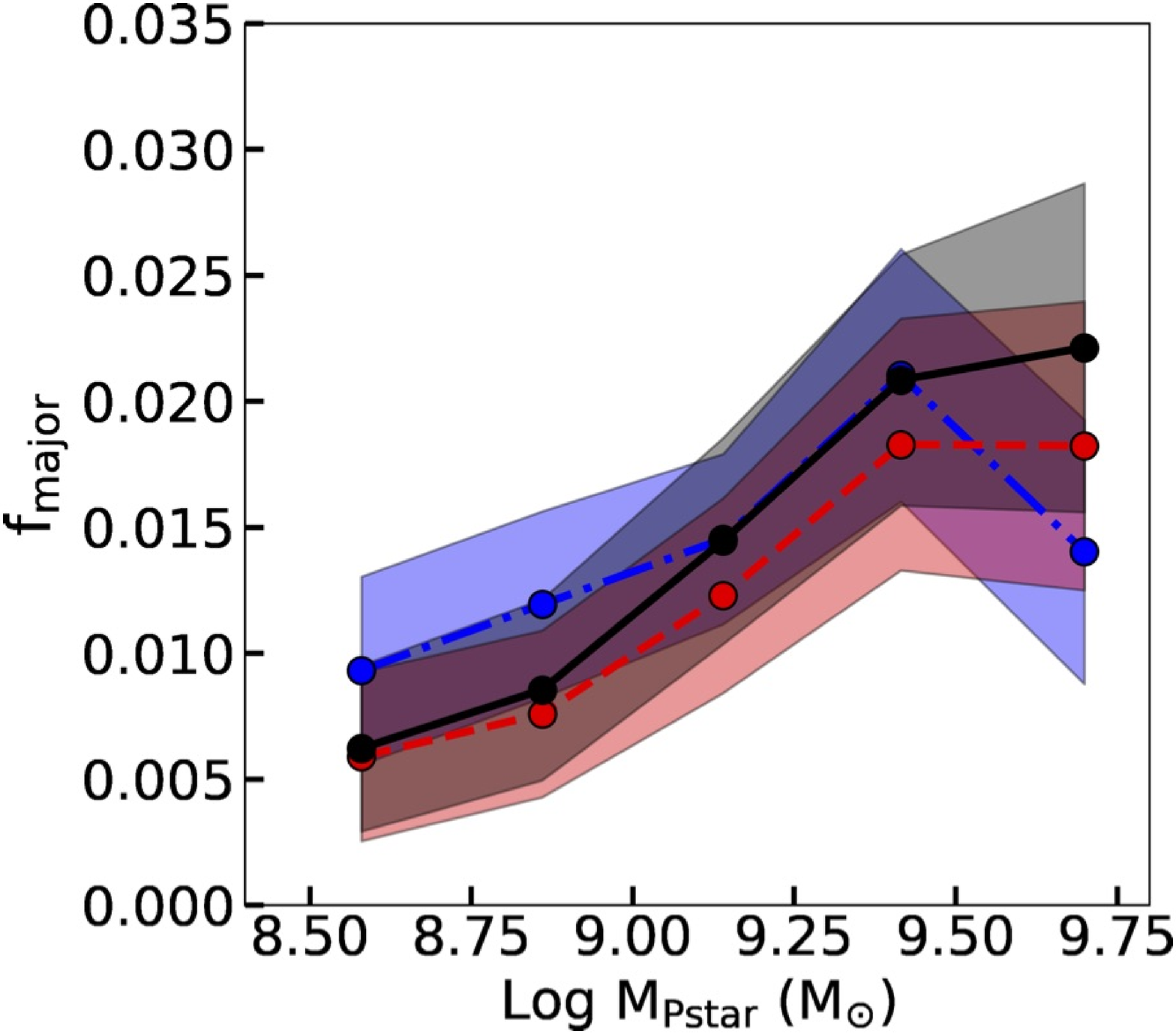}}
 \end{center}
 \caption{\label{fig:DarkVsHydro} 
 {\it Left:} The mean number of companions is plotted per stellar mass bin, $N_{c,m}$. This figure is comparable to Fig.~\ref{fig:Group}.  
 {\it Right:}  The fraction of Primary dwarf galaxies that have a Secondary with a stellar mass ratio of M$_{\rm S,star}/$M$_{\rm P,star} > 1/4$ (``Major Pairs'')
 per Primary stellar mass bin (f$_{\rm major}$). This is comparable to Fig.~\ref{fig:MajMerger}.
 In both panels, mock multiples
  are selected using the {\it Projected} criteria in the {\it Illustris Hydro} catalog (red, dashed) and {\it Illustris-Dark-1} catalog (black, solid line).
 The {\it SDSS} results are plotted in blue (dash-dotted).
Results for both $N_{c,m}$ and f$_{\rm major}$ using the {\it Illustris Dark Projected} catalog are consistently higher than that for {\it Illustris Hydro Projected}, but do agree within 1$\sigma$ (shaded regions) and exhibit the same behaviour as a function of mass.
  } 
\end{figure*}

In the left panel of Fig.~\ref{fig:DarkVsHydro}, we plot the mean number of companions per dwarf per stellar mass bin ($N_{c,m}$) for the {\it SDSS} catalog, {\it Illustris Hydro Projected} catalog and its dark matter only counterpart, marked as {\it Illustris Dark Projected}. This figure is comparable to Fig.~\ref{fig:Group}. Note that the dwarf multiples are selected using only  the {\it Projected} selection criteria of Section~\ref{sec:SelectionP}. The mock catalogs agree within 1$\sigma$, but the {\it Illustris Dark Projected} results are consistently higher. The average $N_c$ for the entire sample of multiples in {\it Illustris Dark Projected} is $N_c = 0.040 \pm 0.005$, in good agreement with the {\it Illustris Hydro Projected} catalog (Table~\ref{table:Nc}).  Results for multiples selected using the {\it Physical} selection criteria show similar agreement.

The right panel of Fig.~\ref{fig:DarkVsHydro} shows the ``Major Pair'' fraction per Primary stellar mass bin for the same catalogs. Again, results agree within 1$\sigma$ and both mock catalogs exhibits the same behaviour as a function of stellar mass. 
We conclude that the results presented in this study using {\it Illustris Hydro} are robust to baryonic effects, which might destroy subhalos or change their kinematics.

\bsp	
\label{lastpage}
\end{document}